# A Model of Gene Expression Based on Random Dynamical Systems Reveals Modularity Properties of Gene Regulatory Networks[†]

Fernando Antoneli[1,4,*], Renata C. Ferreira[3], Marcelo R. S. Briones[2,4]

[1] Departmento de Informática em Saúde, Escola Paulista de Medicina (EPM),
Universidade Federal de São Paulo (UNIFESP), SP, Brasil

[2] Departmento de Microbiologia, Imunologia e Parasitologia, Escola Paulista de Medicina (EPM),
Universidade Federal de São Paulo (UNIFESP), SP, Brasil

[3] College of Medicine, Pennsylvania State University (Hershey), PA, USA

[4] Laboratório de Genômica Evolutiva e Biocomplexidade, EPM, UNIFESP,
Ed. Pesquisas II, Rua Pedro de Toledo 669, CEP 04039-032, São Paulo, Brasil

**Abstract.** Here we propose a new approach to modeling gene expression based on the theory of random dynamical systems (RDS) that provides a general coupling prescription between the nodes of any given regulatory network given the dynamics of each node is modeled by a RDS. The main virtues of this approach are the following: (i) it provides a natural way to obtain arbitrarily large networks by coupling together simple basic pieces, thus revealing the modularity of regulatory networks; (ii) the assumptions about the stochastic processes used in the modeling are fairly general, in the sense that the only requirement is stationarity; (iii) there is a well developed mathematical theory, which is a blend of smooth dynamical systems theory, ergodic theory and stochastic analysis that allows one to extract relevant dynamical and statistical information without solving the system; (iv) one may obtain the classical rate equations form the corresponding stochastic version by averaging the dynamic random variables (small noise limit). It is important to emphasize that unlike the deterministic case, where coupling two equations is a trivial matter, coupling two RDS is non-trivial, specially in our case, where the coupling is performed between a state variable of one gene and the switching stochastic process of another gene and, hence, it is not *a priori* true that the resulting coupled system will satisfy the definition of a random dynamical system. We shall provide the necessary arguments that ensure that our coupling prescription does indeed furnish a coupled regulatory network of random dynamical systems. Finally, the fact that classical rate equations are the small noise limit of our stochastic model ensures that any validation or prediction made on the basis of the classical theory is also a validation or prediction of our model. We illustrate our framework with some simple examples of single-gene system and network motifs.

## 1. Introduction

The importance of chemical fluctuations in gene expression as a potential source of biological information has been well established in the past years. The relevance of what has been called gene expression "noise" for phenotypic diversity in clonal population [1], for cell fate [2] and for several other biological aspects like gene regulation and control [3,4] is broadly recognized. The remarkable difference in the expression of one gene in different cells of the same population (under homogeneous conditions) has been soundly characterized experimentally [5].

---

[†]*Dedicated to our dear friend and colleague Francisco de Assis Ribas Bosco* (1955-2012).

Clearly, the fluctuations observed at the population scale are rooted in smaller scales and the obvious target mechanism to be addressed as the source of fluctuations is the molecular machinery involved in gene regulation. In this perspective, the construction of models that capture the essential dynamical ingredients applicable to each individual gene and are capable to reproduce the statistical properties of ensembles of genes may be very challenging. The main obstacle one must overcome in order to achieve this goal is to understand the basic mechanisms involving a small number of degrees of freedom (possibly related to house-keeping genes) and many regulatory transcription factors acting (simultaneously or not) together with specific enzymes responsible for epigenetic control [6].

The intrinsic stochasticity in gene expression may result from two distinct sources: small number of mRNA and protein molecules, and intermittent gene activity. It is presumed that the first source is the most important in prokaryotic organisms in which the number of mRNA and even protein molecules per cell is very small. In eukaryotic organisms, and especially in higher eukaryotic organisms, where the number of proteins is fairly large, the main source of stochasticity is the intermittent activation of the gene. Typically, in order to activate an eukaryotic gene, several transcription factors are needed together with chromatin remodeling, and therefore one expects to observe longer periods of gene activity and inactivity resulting in large bursts of mRNA molecules and proteins.

Based on clear experimental results, theoretical efforts have been made in the last ten years aiming at finding models to describe gene activity in time. These efforts were driven by the desire to capture the essential aspects of gene expression fluctuations and to explain the phenomenon at the population scale [7,8]. Different types of models and approaches have been proposed from the most appealing ones based on Langevin-like equations [9–11] (see also [12] and references therein) to the evolution of densities based on master equations [13–20] in analogy with the quantum many-body problem [21]. In general, those models assume that the underlying stochastic process has some specific property like stationarity or is of Markov type. For instance, the Langevin approach assumes that the stochastic component is given by additive white noise, a formulation especially adequate for computational purposes [22,23].

Since genes do not work in isolation, the investigation also must include the interactions between the various elements involved in the transcription and translation process, the so called *transcription regulatory network*. Genes can be viewed as "nodes" in the network, with inputs being proteins such as *transcription factors*, and outputs being the level of gene expression. These regulatory networks are only beginning to be understood, and one of the outreaching aims of *systems biology* is to model their behavior and discover their *emergent properties*.

The theory of networks and its application in several branches of science such as physics, biology, sociology, just to name a few, has attracted a lot of interest in the last thirty years [24]. Much progress has been made on "static" aspects of networks, i.e., how the combinatorial features of a network affect their topological or statistical behavior. The work of Watts and Strogatz [25] on "small world" networks is especially well known. Probabilistic aspects of networks have also received a lot of attention, see for instance [26,27]. However, general studies of the *dynamics of networks*, that is, the *processes* that take place on a network are more scarce. Many papers on network dynamics make restrictive assumptions – a typical one being to assume weak linear coupling. There are also several papers about numerical simulations of specific models. In an attempt to revert this situation, two formal – and essentially equivalent – frameworks have been proposed for the *deterministic nonlinear dynamics* of networks: one based on "groupoids" by [28,29] and one based on the combinatorics of couplings by [30]; see also [31] for a comprehensive review of the "groupoid formalism" and [32] for an up to date presentation of the combinatorial formalism of [30] and a comparison between both approaches. One of the main features of these approaches is the possibility to formulate and prove generic dynamical properties associated with the global topology of coupled cell networks such as: the existence of robust patterns synchrony supporting stationary and periodic solutions in regular networks [33], canard cycles and their explosion [34], existence of robust heteroclinic cycles [32], etc. So far, both frameworks have been developed to study networks whose architecture are fixed and whose dynamics is described by deterministic equations.

In order to investigate the dynamics of regulatory gene networks on a general basis we shall propose a framework, inspired by the groupoid formalism, for building *stochastic dynamical models* associated to a given *arbitrary network topology*. We shall adopt the standpoint that a group of clonal cells submitted to the same environmental conditions represent a biological domain where the set of copies of a given gene (no more than one copy per cell) forms the statistical ensemble that contains all the relevant dynamical and statistical information about the gene in the given conditions. Distributions over this ensemble characterize properties of the cell population relative to the gene under consideration. The framework introduced here is motivated by eukaryotic unicellular organisms, the main archetype being *Saccharomyces cerevisiae*, whose genetic regulation mechanisms have been extensively studied [35]. Even so, our framework can also be appropriately interpreted as a model for prokariotic cells.

One of the virtues of the approach described in this paper is that it allows to overcome the difficulties encountered in the master equations and Langevin equation formalisms when one tries to extend the theory from single-genes to networks.

In the first case it is not possible to couple two master equations in such a way that the coupling represents an interaction between the corresponding physical variables, as it is usually done with ODE's, since a master equation is a differential equation for the probabilities and not to the corresponding physical variables. For instance, it is the actual number of mRNA's that influence the production of proteins not the probability of production of one or more mRNA molecules. In the second case, the variables in the equations do represent physical quantities but there are extremely hard technical problems in treating the white noise variable that drives the Langevin equation. When the dependence of the equation on the white noise is linear, the theory of stochastic integrals provides a solid ground for a rigorous treatment of this type of equation. However, this property of linearity only holds in very restricted types of coupling (additive coupling), which barely embodies the most general circumstances.

On the other hand our framework is much more flexible in the sense that it provides a general prescription to couple the "single gene systems" in a way that is consistent with *any* given network topology. The "single gene systems" should form an *admissible class* of building blocks for the *internal dynamics* of a regulatory gene network, that is, they must provide a model for the regulation, transcription and translation processes of a single gene. This approach is inspired by the "groupoid formalism" for network dynamics [31], where the dynamics of the network is defined first by fixing the internal dynamics of each node and then coupling these internal dynamics according to the topology of the network.

The prototype for the dynamics of individual genes considered here may be viewed as the stochastic version of the *classical rate equations*, as advanced in [36]. More precisely, we consider the single gene dynamics of [36] in a simplified and time-discretized form that takes into account only two effects: the degradation (decay) and stochastic activation (or repression) by transcription factors – the switching process. This naturally leads to the class of affine random dynamical systems as the basic internal dynamics, which is the minimal class of RDS that is admissible and, at the same time, amenable to mathematical analysis [37,38]. As it turns out, a coupled network formed by affine random dynamical systems obtained by our method of coupling is again an affine random dynamical system. The typical switching process generated by the action of the transcription factors is a discrete-time stationary stochastic process and therefore the analysis of the single-gene model only requires basic theory of stochastic processes at the level of [39]. Finally, it is worth to remark that a continuous-time version of the single gene dynamics obtained here has been studied in [40–43].

**Structure of the paper.** The paper is organized as follows: in Section 2.1 we explain our framework by defining a class of internal dynamics (also called single-gene systems) and state the main results. In Section 2.2 we discuss some typical single-gene systems and determine their statistical properties as illustrative examples (these examples are not necessary for applications of the general theory). In Section 3 we present our conclusions based on the theory introduced previously. In Sections 4.1 – 4.4 we present the mathematical background and the main arguments underpinning our theory. In Sections 4.5 – 4.9 we determine the statistical properties of single gene systems. This last part is independent of the remaining parts of the paper and is not necessary for the applications of the theory expounded here.

## 2. Results and Discussion

In the first section we define our basic single-gene model for the internal dynamics and state our main results. In the second section we discuss some typical examples of single-gene models.

### 2.1 The General Formalism

Before stating our main results let us precisely define the basic building blocks of our formalism: the 2-dimensional affine random dynamical system. The state variables $(x_n, y_n)$ represent the two quantities of interest: $x_n$ is the number density of mRNA molecules produced between two successive observations and $y_n$ is the number density of protein molecules produced between two successive observations. The dynamics is given by

$$\begin{aligned} x_{n+1} &= (1-\gamma)x_n + \delta \xi_n \\ y_{n+1} &= (1-\alpha)y_n + \beta x_n \end{aligned} \quad (*)$$

where the mRNA production rate $\delta$, the mRNA degradation rate $\gamma$, the protein production rate $\beta$ and the protein degradation rate $\alpha$ are constant in time and $\xi_n$ is a discrete time finite state stochastic process – the switching process. The phase space of system (*) is the rectangle $[0,\delta/\gamma] \times [0,\beta\delta/(\alpha\gamma)]$ (here one must assume that $0 < \alpha, \beta, \gamma, \delta < 1$). The switching process $\xi_n$ is usually a binary process with states {0=OFF, 1=ON}, which is assumed to be stationary. Finally, we suppose that the initial conditions $x_0$ and $y_0$ are mutually independent random variables and are independent from the switching process $\{\xi_n\}$.

Given a Gene 1 with state variables $(u_n, v_n)$, a Gene 2 with state variables $(x_n, y_n)$, both represented by systems of the for (*), and a link from Gene 1 to Gene 2, the coupling between the two corresponding systems of is provided by an *input function*, that is, a continuous strictly increasing function $p:[0,\varepsilon] \to [0,1]$ defined on the range of the protein variable $v_n$ of Gene 1, where $\varepsilon = \beta\delta/(\alpha\gamma)$ is the maximum value of the protein number density. The input function is used to define the probability distribution of the switching process $\xi_n$ of Gene 2 in the following way: $p_{ON} = \mathbf{P}(\xi[v_n]=1) = p(v_n)$ is the probability that transcription of Gene 2 is turned ON, at time $n$, due to the presence of the transcription factor produced by Gene 1 acting at the promoter, at time $n$, and $p_{OFF} = \mathbf{P}(\xi[v_n]=0) = 1 - p(v_n)$ is the probability that transcription is turned OFF due to the absence of transcription factor produced by Gene 1 acting at the promoter, at time $n$. In order to model the situation where the transcription factor acts by *repression* at the promoter region, instead of *activation*, one simply employs the function $q(v_n) = p(\varepsilon - v_n)$. Note that both probabilities $p_{ON}$ and $p_{OFF}$ depend on the state variable $v_n$ denoting the concentration of the upstream transcription factor here represented by Gene 2.

The most prominent examples of input functions is comprised by the extensively used family of *Hill functions* $H(r,k,y) = y^r/(k+y^r)$, where $k$ is called the *dissociation constant* and $r$ is called *Hill coefficient* [44]. They are frequently employed to describe the proportion of molecules saturated by a ligand as a function of the ligand concentration; they are also used in determining the degree of cooperativeness of the ligand binding to an enzyme or receptor. The dissociation constant is such that when $y = k^{1/r}$ the function $H$ reach half of its maximum value: $H(r,k,k^{1/r}) = 1/2$. The Hill coefficient $r$ represents the *degree of cooperativity* of the ligand that is binding to the receptor. It also determines the *sigmoidicity* of the curve $H$ versus $y$: the larger $r$, the steeper the curve near $y = k^{1/r}$. Values of $r > 1$ denote positive cooperativity, while values $0 < r < 1$ denote negative cooperativity. The value $r = 1$ denotes completely independent binding, regardless of how many additional ligands are already bound, this is the well known *Michaelis-Menten kinetics*. When $r = 0$ the Hill function is constant $H(0,k,y) = 1/(k+1)$ and thus there is no self-regulation anymore. Finally, when $r \to +\infty$, the Hill function converges to the *Heavisde function* $H_1(k,y) = \{0$ if $y < 1$; $1$ if $y > 1$; $1/(k+1)$ if $y = 1\}$ and $k^{1/r} \to 1$. This limit function provides a deterministic binary switching between the states ON and OFF with jump at $y=1$.

**Theorem A.** *Let G be a directed graph representing the topology of a network. Let be given the type of each link (activator or repressor) and the specification of the logical gating between the links entering nodes with multiple inputs. Assume that each node of G is attached to a single-gene model, given by 2-dimensional affine random dynamical system of the type (\*), representing the internal dynamics of the corresponding gene. Then the explicit prescription in terms of input functions, as given above, for coupling the internal dynamics, provides a fully coupled stochastic gene regulatory network corresponding to the topology G which is an affine random dynamical system of dimension twice the number of nodes of G (see Sections 4.1– 4.3 for details).*

Theorem A would be trivial in the deterministic context, since coupling two deterministic equations is a simple matter. However, coupling random dynamical systems is not as straightforward, specially in our case where the coupling is between a state variable of one gene and the switching stochastic process of another gene. Indeed, it requires a non-trivial argument to show that the composed system formed by coupling several 2-dimensional single gene random dynamical systems is again a random dynamical system.

A fundamental consequence can be drawn from Theorem A by the application of the general theory of random dynamical systems [38,45], namely, a stochastic regulatory network defined be affine random dynamical systems as in theorem A has a unique *stationary distribution*. This follows from the fact that the full system given by Theorem A is an affine RDS and the fact that $\max\{(1-\gamma),(1-\alpha)\} < 1$ (see Section 5.6, page 221 of [38]).

Our next result concerns the average equations provided by our stochastic model: the classical *rate equations of kinetic theory* can be obtained from the stochastic model presented here, by taking average of the state variables.

**Theorem B.** *The classical rate equations of kinetic theory coupled through Hill input functions can be readily obtained from the stochastic model provided by theorem A, by taking averages of the state variables in the internal dynamics of each single gene sub-system (see Section 4.4 for the details).*

Therefore, any validation or prediction obtained through a deterministic model also holds "on average" for the stochastic model presented here. In particular, the whole dynamical theory developed in [46], which is completely based on deterministic rate equations can be reproduced by our formalism in the small noise limit.

Theorem A corroborates the view that the "dynamical modularity" of gene regulatory networks holds at the stochastic level, as well as it holds at the deterministic level. Theorem B establishes that these two dynamical frameworks are related as expected, namely, the deterministic counterpart emerges as "small noise limit" of the stochastic model.

By *dynamical modularity*, we mean the fact that the dynamics of such networks can be built from certain elementary building blocks represented by the nodes of the network and the links between these nodes represent the couplings between the elementary building blocks. Observe that this form of modularity does not implies reductionism, since the coupling functions are usually non-linear and therefore the collective dynamics of the network can not be deduced by studying its components individually.

Notwithstanding its relation with the deterministic theory, the stochastic framework introduced here does provide new features to dynamics of gene regulation. For instance, the bifurcation theory of random dynamical systems is very subtle and still poorly understood, even in the most simple cases [47,48] – despite the fact that the corresponding deterministic bifurcation theory have been completely established. These phenomena may arise in our framework, since the class of models used for the internal dynamics can be extended, without any extra effort, to include some non-linear systems (affine RDS forms the minimal class of admissible internal dynamics), in such a way that the main results of the theory (theorems A and B) continue to hold.

*2.2  Examples: Single gene systems*

We shall briefly examine some simple examples of single gene systems, some of them have been exhaustively studied and do occur in real regulatory networks, see [35] – also called *externally regulated genes* are the "fundamental building block of gene regulatory circuits" [49]: (i) the single gene with IID and (ii) the single gene with Markovian switching. The assumption that the gene is "externally regulated" means that the distribution probability of the switching process is constant and depends on transcription factors that are not part of the regulatory network and have (approximately) constant concentration.

Before starting the analysis let us recall some terminology. It has become customary in the literature on gene expression to quantify the *mean expression level* of mRNA or protein of gene by the mean value of the corresponding random variable and to quantify the amount of spreading of the expression level, also called *fluctuation of the expression level* or *noise level*, by a normalized measure of dispersion. The most popular measures of dispersion used are the *relative standard deviation* or *coefficient of variation* $\eta^2 = \sigma^2/\mu^2$ and the *index of dispersion*, also known as *Fano factor*, $D = \sigma^2/\mu$, where $\mu$ is the mean and $\sigma^2$ is the variance and $\eta^2 = D/\mu$.

**Single gene with IID switching process.** This is the simplest possible single gene system with the switching process $\xi_n$ given by a sequence of independent and identical binary Bernoulli processes with (constant) probability $p$ of turning ON the transcription of the gene – thus the mean value of $\xi_n$ is $p$ and the variance is $p(1 − p)$. In this case, the expression levels of mRNA and of protein in a gene with (approximately) constant rates of production and degradation essentially depend on the probability $p$ of turning the gene ON.

A single gene system with IID switching and constant reaction rates $\alpha$, $\beta$, $\gamma$, $\delta$ is given by equation (*), where $\xi_n$ is a Bernoulli process with probability $p$. It is a stochastic Markovian process with exponential decay of correlations at a rate $\lambda \approx -\log[\max\{(1-\gamma),(1-\alpha)\}]$. At equilibrium, the mean expression levels of mRNA and protein are, respectively, $\mu_x = \delta/\gamma\, \mu_\xi$ and $\mu_y = (\beta\delta)/(\alpha\gamma)\, \mu_\xi$. The coefficient of variation of the switching process is $\eta_\xi^2 = (1 − p)/p$ and the fluctuations of the expression levels of mRNA and protein are, respectively, $\eta_x^2 = \eta_\xi^2\, \gamma^2/(1-(1-\gamma)^2) \approx \gamma\, \eta_\xi^2$ and $\eta_y^2 = \eta_\xi^2\, (\alpha\gamma)^2/[(1-(1-\gamma)^2)(1-(1-\alpha)^2)][(1+(1-\gamma)(1-\alpha))/(1-(1-\gamma)(1-\alpha))] \approx \gamma\alpha\, \eta_\xi^2$. In particular, we have that $\eta_x^2 \approx \delta(1-p)/\mu_x$ and $\eta_y^2 \approx \beta\delta\,(1- p)/\mu_y$. See Sections 4.5 and 4.6 for details.

This result shows that the expression level and the corresponding noise level of a gene with (approximately) constant rates of production and degradation comes essentially form the switching mechanism taking place at the promoter region of the gene. As expected, the mean expression levels agree with the classical deterministic result (equilibrium values).

It is interesting to compare this result with the corresponding result obtained with the master equation formalism. The standard calculation of the fluctuation of expression level using the master equations formalism is presented in [8], for instance. The result for the noise level of $\xi$ in obtained above is essentially (up to a multiplicative normalizing factor) the same as in equation (6) of [8], while in the mRNA and protein cases, our model reproduces the so called "Binomial" part of equations (4) and (5) of [8]. This discrepancy happens because a model of gene expression based on master equations is essentially a continuous-time birth and death processes driven by a Markovian switching [13]. In other words, transcription and translation are typically assumed to follow Poisson processes where the production probabilities per time unit are proportional to the number of active genes and mRNAs, respectively. This is an implicit modeling assumption about the real process that is deeply rooted into the underlying formalism, hence the additional "Poissonian" terms appearing in [8] are an inescapable feature of any description based on master equations.

**Single gene with Markovian switching.** Let $\xi_n$ by a binary Markov chain determined by a two-by-two stochastic matrix $P$ with off-diagonal elements $0 < p, q < 1$ and diagonal elements $(1−p)$ and $(1−q)$, with $q \neq 1−p$ (the equality gives the IID case). Here, $p$ is the probability of transition from state OFF to state ON and $q$ is the probability of transition from state ON to state OFF.

The stationary distribution of $\xi_n$ is given by the row vector $\pi = (\pi_{OFF}, \pi_{ON}) = (q/(p+q), p/(p+q))$, where $\pi_{OFF} = (1-\pi_{ON})$. Therefore, at equilibrium, the expectation value of $\xi_n$ is $\mu_\xi = \pi_{ON} = p/(p+q)$, the variance is $\sigma^2_\xi = pq/(p+q)^2$, the dispersion index is $D_\xi = \pi_{OFF} = q/(p+q)$ and the coefficient of variation is $\eta_\xi^2 = (1-\pi_{ON})/\pi_{ON} = q/p$. The temporal auto-correlation function is $\rho_\xi(n) = (1-p-q)^{|n|}$ for all $n \neq 0$ and $\rho_\xi(0) = \sigma^2_\xi = pq/(p+q)^2$.

A single gene system with Markovian switching and constant reaction rates $\alpha, \beta, \gamma, \delta$ is given by equation (*), where $\xi_n$ is a binary Markov chain determined by a two-by-two stochastic matrix $P$ with off-diagonal elements $0 < p, q < 1$. At equilibrium, the mean expression levels of mRNA and protein are, respectively, $\mu_x = \delta/\gamma \, \mu_\xi$ and $\mu_y = (\beta\delta)/(\alpha\gamma) \, \mu_\xi$. The fluctuation of expression levels of mRNA and protein are, up to first order in auto-correlation, given by $\eta_x^2 \approx \gamma \, \eta_\xi^2 \, (1+2(1-\gamma) \, \rho_\xi(1)/\rho_\xi(0))$ and $\eta_y^2 \approx \gamma\alpha \, \eta_\xi^2 (1+2((1-\gamma)+(1-\alpha)(1-(1-\gamma)^2)) \, \rho_\xi(1)/\rho_\xi(0))$. In particular, we have that $\eta_x^2$ is proportional to $\delta(1-\pi_{ON})/\mu_x$ and $\eta_y^2$ is proportional to $\beta\delta(1-\pi_{ON})/\mu_y$. See Sections 4.5 and 4.7 for details.

This result shows, as in the IID case, that the expression level of a gene with (approximately) constant rates of production and degradation comes essentially form the switching mechanism taking place at the promoter region of the gene. However, the fluctuations of the expression level not only depend on the noise level but also depend on the temporal auto-correlation function of the switching process. Again, as in the IID case there is a term corresponding to the noise level of the switching process which is the "Binomial" part of the equation. The partial agreement between expressions coming from distinct approaches suggest that the concurring parts represent "model-independent" characteristics of gene expression, while the differences are related to the specifics of the underlying modeling (explicit or implicit) assumptions.

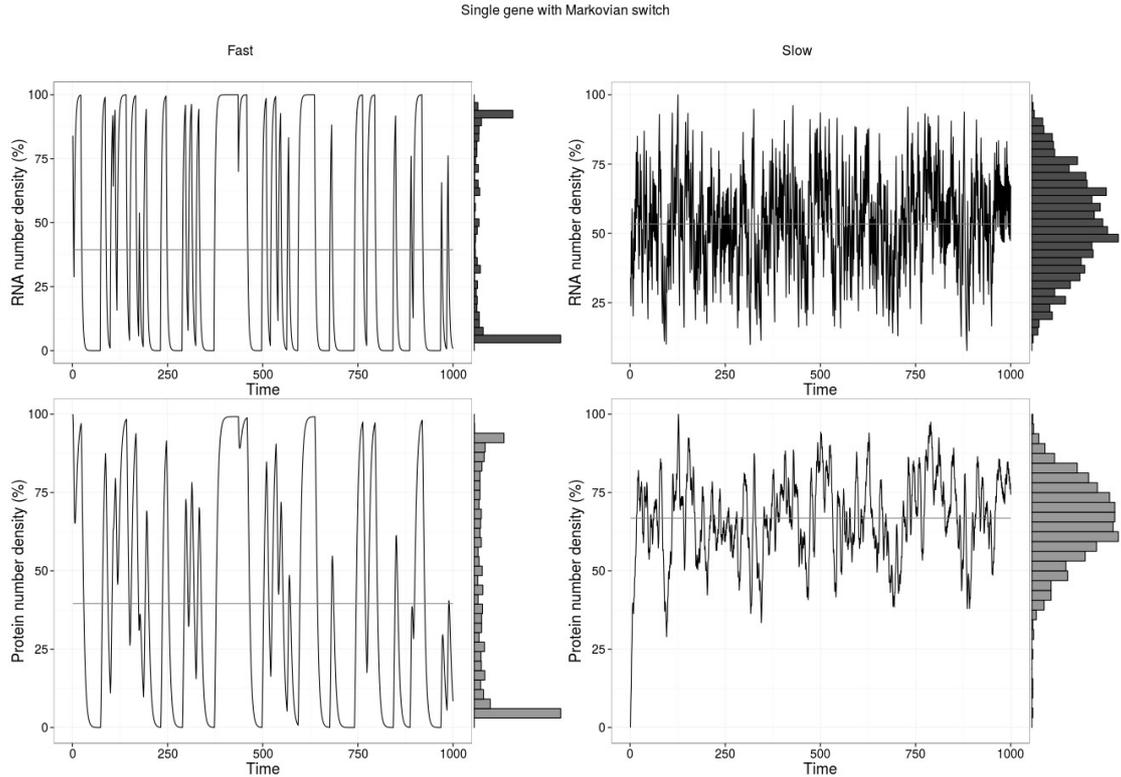

**Figure 1.** Simulation of typical time series and stationary distributions of a single gene with Markovian switch, with "fast switch" (left) and "slow switch" (right). The time scale (abscissa axis) is the average time to transcribe 1 gene ($\approx$ ½ min in yeast). The number density scale (ordinate axis) is the relative concentration of molecules (%) with respect to its maximum value. The horizontal line is the mean value of the level of expression.

The common "Binomial" terms appearing in our formalism and the master equation formalism are related to the binary character of the switching mechanism, while additional "Poissonain" terms of the master equation approach reflect the Poissonian character inherent to the underlying formalism.

By comparing our two examples, one easily sees that both switching mechanisms are able to recover the fundamental relations $\eta_x \sim 1/\sqrt{\mu_x}$ and $\eta_y \sim 1/\sqrt{\mu_y}$, that is, the mRNA/protein noise scales with the mRNA/protein level as the inverse square root. However, there is a crucial difference between the IID and the Markovian switching. While in the former there is only one phase, in the latter it is possible to distinguish two very distinct phases: the "fast switch" and the "slow switch", due to the fact that in the Markovian case there are two transition probabilities for turning on and turning off and they both can be set independently – in the IID case the probabilities are complementary. The fast switch is characterized by small values of the transition probabilities (high rate) of the switching process and displays bimodal distributions for the mRNA and protein density numbers while the slow switch is characterized by larger values of the transition probabilities (low rate) unimodal distributions (see Fig. 1).

In any a single gene model one observes that the relative strength of stochastic effects grows as the number of reacting molecules decreases, and thus one expects that the stochasticity due to switching of the gene status is the most important, at least for eukaryotic organisms. Accordingly, one may neglect the mRNA/protein production and decay rates randomness. In our model this is readily incorporated in the equations by assuming that the reaction rates are constant in time.

### 2.3 Examples: Network motifs

As mentioned before, in a single gene model, it is common to assume for simplicity that gene activation or repression is due to a single molecule; however, a gene inside a regulatory network is turned ON (or OFF) due to the collective action of several different regulatory factors. In the next three examples we show how to implement multiple gene regulation within our framework: (i) the auto-regulated motif; (ii) the two-component loop and (iii) the feed-forward loop. Although a full analysis network motifs is not our main concern in this paper, these systems yield representative examples to illustrate our formalism and are enough to give the reader a "guidebook" on how apply it to *any* network topology.

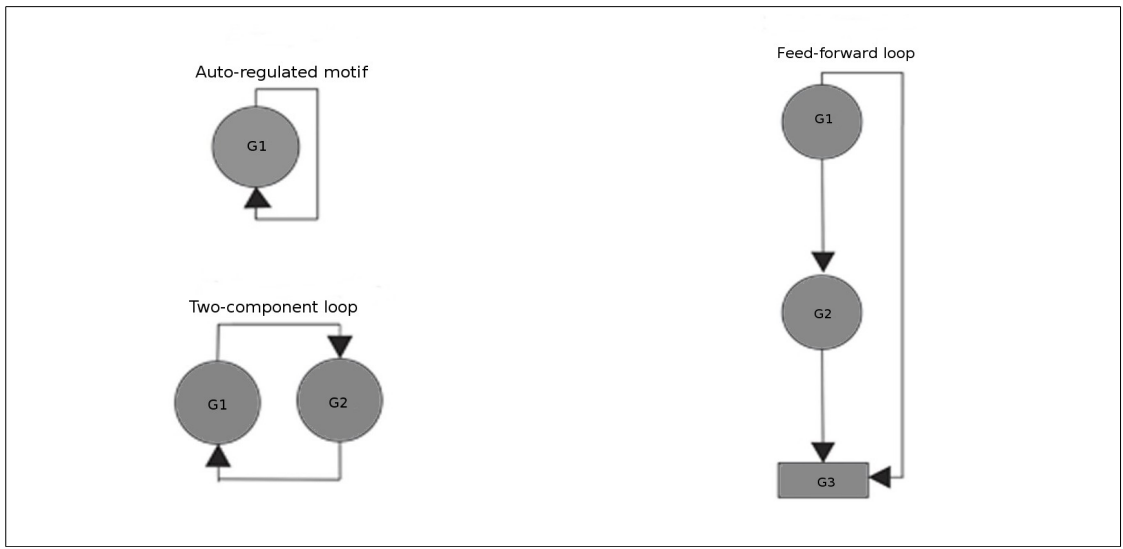

**Figure 2.** Most common network motifs found the regulatory network of *Saccharomyces cerevisiae* [35]. The auto-regulated motif (top-left), the two-component loop with two genes (bottom-left) and the feed-forward loop (right). Here we have represented only the network topology and the orientation of the links.

**Auto-regulated motif.** This is the smallest possible network motif, also called "self-regulated gene", composed by only one gene (see Fig. 2 (top right)). In this system a single gene is coupled to itself in such way that the probability distribution of the switching process explicitly depends on the protein number density produced by the gene itself. More formally, an auto-regulated motif is given equation (*) with constant reaction rates $\alpha$, $\beta$, $\gamma$, $\delta$ and transcription switching given by a stochastic process $\xi_n = \xi[y_n]$, whose probability distribution depends on the protein number density through an input function $p:[0,\varepsilon] \to [0,1]$ defined on the range $[0,\varepsilon]$ of the protein variable, where $\varepsilon = \beta\delta/(\alpha\gamma)$ is its maximum value.

If we assume that $p(y) = H(k,r,y)$ is a Hill function and that $\varepsilon > 1$, then the mean expression level of the protein $\mu_y$ is approximately given by a solution of the *fixed-point equation* $h(y) = y$, where $h(y) = \varepsilon H(k,r,y)$. In the repressor case, there is typically only one solution and, when $r$ is large, one has that $\mu_y \approx \varepsilon - 1$. In the activator case, there are typically three solutions and, when $r$ is large, one has that $\mu_y \approx \varepsilon/2$ and the probability distributions of the mRNA and the protein are given by a mixture of two unimodal distributions, one with mode near 0 and the other with mode near $\varepsilon$. Finally, the fluctuations of expression levels of the protein are $\eta_y^2 \approx 1 - 1/\varepsilon^4\ O(1/r^4)$ in the repressor case and $\eta_y^2 \approx \varepsilon^2\ O(1/r^4)$ in the activator case. The fluctuation of expression level of the protein shows that self-activation leads to *over-dispersed* noise ($\eta_y^2 > 1$) and self-repression leads to *under-dispersed* noise ($\eta_y^2 < 1$); when there is no self-regulation ($r = 0$) one is back to the situation of the single-gene with IID switching (with $p = 1/(1+k)$).

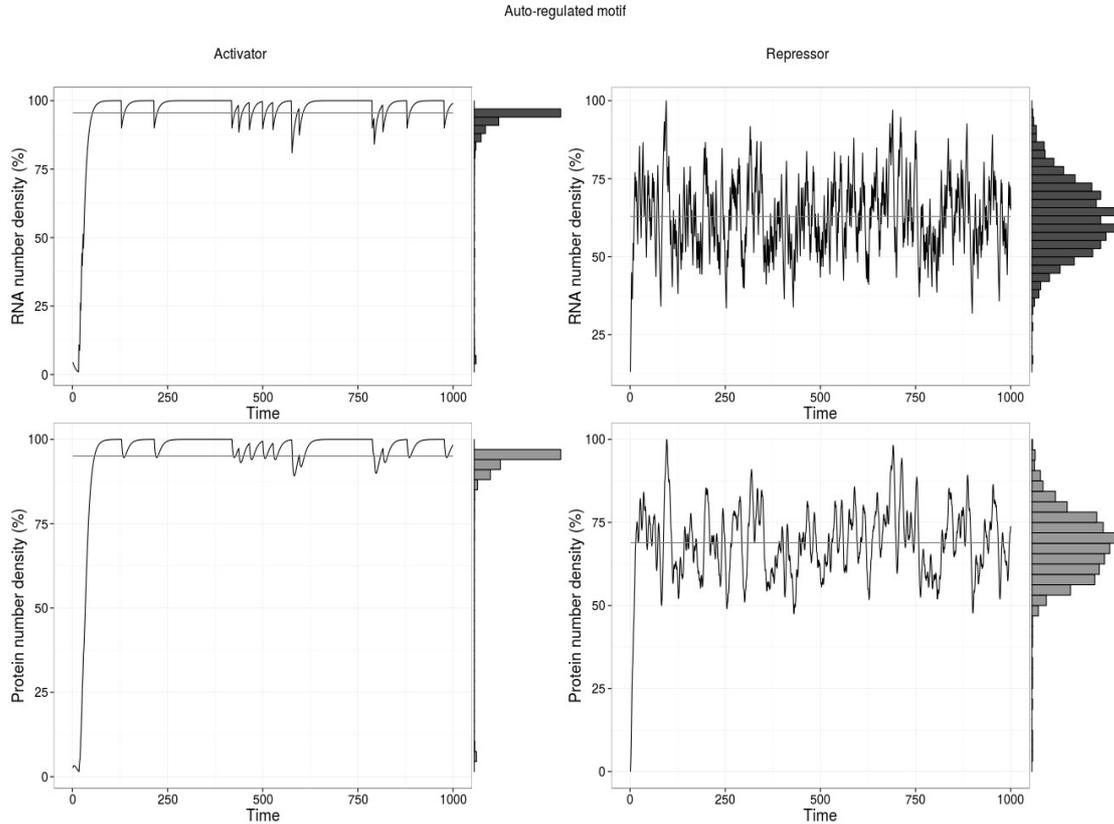

**Figure 3.** Simulation of typical time series and stationary distributions of an auto-regulated motif in the activator case (left) and the repressor case (right). The time scale (abscissa axis) is of the order of the average time to transcribe 1 gene ($\approx$ ½ min in yeast). The number density scale (ordinate axis) is the relative concentration of molecules (%) with respect to its maximum value. The horizontal line is the mean value of the level of expression. Note that in the activator case, the stationary distribution produced by the single time series is concentrated at the full production mode. The parameters were chosen to reproduce the qualitative behavior and do not represent real values in any specific gene.

The self-activator with three fixed points displays a very high ratio of the variance to the mean, which is not very informative about the fluctuation of expression level because the distribution is bimodal. However, this bimodality is different from the one observed in the Markovian gene. Here the bimodality is due to the bistability of the system: the self-activator behaves in an almost binary fashion, having a threshold production value, below which the production is next to null and above which the production is almost full. Thus, the bimodality may be seen only on the distribution obtained from an ensemble, and not in the distribution computed from only one time series, which will be concentrated at one of the modes (see Fig. 3).

The self-repressor, in contrast, has unimodal distribution, and the ratio of the variance to the mean is always a good description of the fluctuation of the expression level. These results agree with the observations of [49] for the master equation of a self-repressor and of a self-activator (see figure 4(b) and 4(c) of [49]). The self-repressor has a rather steady self-sustained production level near the expected level of expression, which is a very important feature to achieve homeostasis (see Fig. 3).

The characteristics of auto-regulated motifs come from their strong feedback mechanism and cause them to rarely appear in isolation in gene regulatory networks and in fact, in order to be useful, auto-regulated motifs need some sort of "external control" in the form of one or more additional transcription factors acting as "control levers". Typically, a gene regulatory networks contains a dozen auto-regulated motifs which are never isolated, but highly wired into the network – it has been reported that there are at least 10 auto-regulated motifs in the regulatory network of *Saccharomyces cerevisiae*, whose interaction have been confirmed by traditional chromatin immuno-precipitation, among the 106 analyzed transcription factors [35].

**Two-component loop.** The simplest network motif of this type is the two-component loop, a network with two nodes {Gene 1, Gene 2} and two undirected links between these nodes (see Fig. 2 (bottom right)). It has been reported that there are at least 3 two-component loops in the regulatory network of *Saccharomyces cerevisiae* [35].

There are 3 possible combinations for the action of the two transcription factors: repressor-repressor, activator-activator, activator-repressor. The system of affine RDS corresponding to the two-component loop is the following:

$$\begin{aligned} u_{n+1} &= (1-\gamma_1) u_n + \delta_1 \zeta[y_n] \\ v_{n+1} &= (1-\alpha_1) v_n + \beta_1 u_n \\ x_{n+1} &= (1-\gamma_2) x_n + \delta_2 \xi[v_n] \\ y_{n+1} &= (1-\alpha_2) y_n + \beta_2 x_n \end{aligned} \qquad (**)$$

where $(u_n, v_n)$ are the state variables of Gene 1 and $(x_n, y_n)$ are the state variables of Gene 2. The reaction rates of Gene 1 are $\alpha_1, \beta_1, \gamma_1, \delta_1$, the reaction rates of Gene 2 are $\alpha_2, \beta_2, \gamma_2, \delta_2$. The couplings are given by two input functions $p_{12}(y) = H(k_1, r_1, y)$ and $q_{21}(v) = H(k_2, r_2, v)$, where $p_{ON} = \mathbf{P}(\zeta[y_n] = 1) = p_{12}(y_n)$ (or $p_{12}(\varepsilon_2 - y_n)$) and $q_{ON} = \mathbf{P}(\xi[v_n] = 1) = q_{21}(v_n)$ (or $q_{21}(\varepsilon_1 - v_n)$).

The two-component loop displays a certain amount of feedback but not as strong as in the auto-regulated motif. In the case with two mutually repressing genes the system displays a bimodal stationary distribution for the protein number density of both genes, leading to a shallow valley between the two peaks, as reported in [49] (see Figure 4). Even though the bimodality in this case is much softer than the one encountered in the self-activator the mean expression level stays very near the valley and hence the ratio of the variance to the mean is not very informative about the fluctuation of expression level, as well.

It has been observed in [50], using the deterministic framework, that a two-component loop with two mutually repressing genes (also called "double negative feedback") exhibits two stable steady states in a certain parameter range and is able to function as toggle switch. The stochastic version introduced here also exhibits bistability when the Hill coefficients are high enough ($r_1, r_2 > 1$) – with the advantage that the switching between the two stable states is naturally induced by the noise – in this case, as in the self-activator, bimodality may be seen only on the ensemble distribution.

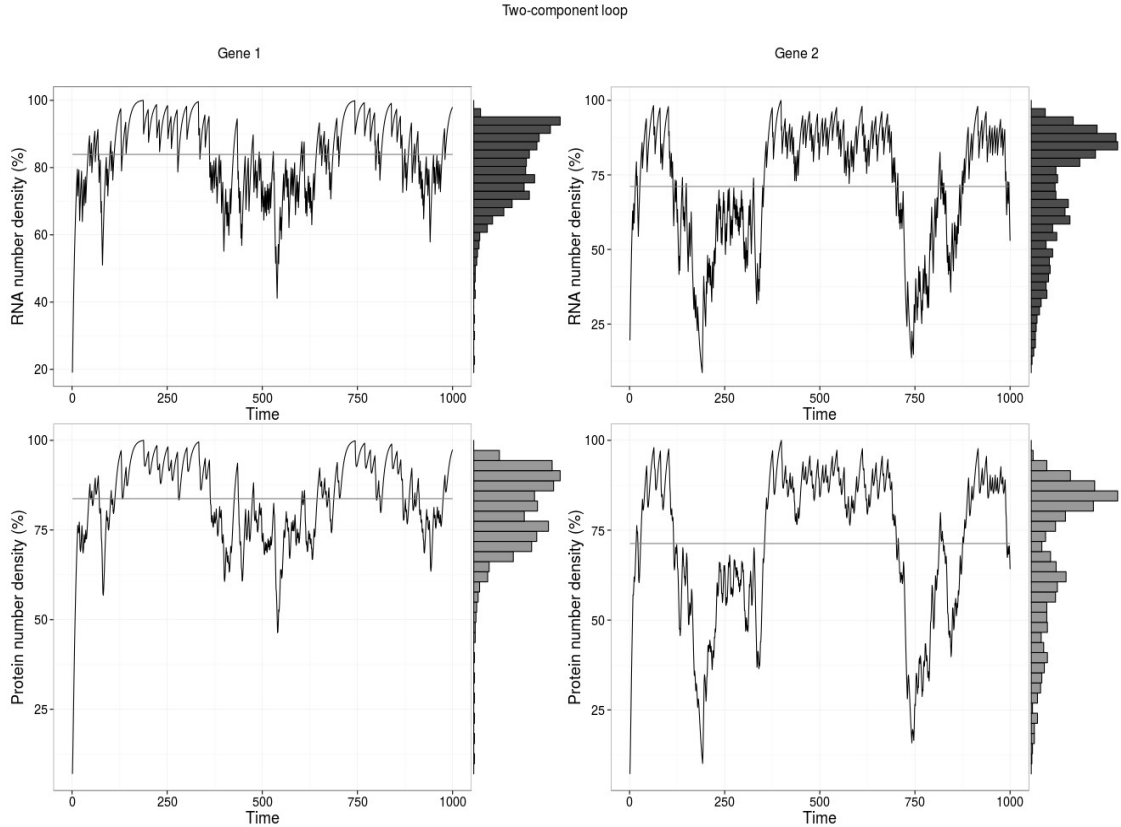

**Figure 4.** Simulation of typical time series and stationary distributions of a two-component loop in the repressor-repressor case, Gene 1 (left) and Gene 2 (right). The time scale (abscissa axis) is of the order of the average time to transcribe 1 gene ($\approx$ ½ min in yeast). The number density scale (ordinate axis) is the relative concentration of molecules (%) with respect to its maximum value. The horizontal line is the mean value of the level of expression.

**Feed-forward loop.** The ubiquitous *feed-forward loop* (FFL) is a network motif with 3 nodes {Gene 1, Gene 2, Gene 3}, one link from Gene 1 to Gene 2, one link from Gene 1 to Gene 3 and one link from Gene 2 to Gene 3 (see Fig. 2 (left)). In addition, Gene 1 is assumed to be externally regulated by an IID or a Markovian switching. It has been reported that there are at least 49 feed-forward loops, potentially controlling 240 genes, in the regulatory network of *Saccharomyces cerevisiae*, among the 106 analyzed transcription factors [35].

According to whether each link represents an activator or a repressor, there are $2^3 = 8$ possibilities for a feed-forward loop. These 8 possible feed-forward loops can be classified into 2 types: coherent and incoherent. This classification is based on the attribution of a "sign" to each link in the network. Activation is associated with positive control, hence an activator link is given the plus sign (+). On the other hand, repression is associated with negative control, hence a repressor link is given the minus sign (−). The *overall sign* of a sequence of two consecutive links is given by the multiplication of the sign of each link (so that two minus signs give an overall plus sign). The *coherent feed-forward loop s* are the ones where the link from Gene 1 to Gene 3 has the *same* sign as the overall sign obtained from the sequence formed by the links from Gene 1 to Gene 2 and from Gene 2 to Gene 3. The *incoherent feed-forward loops* are the ones where the link from Gene 1 to Gene 3 has the *opposite* sign as the overall sign obtained from the sequence formed by the links from Gene 1 to Gene 2 and from Gene 2 to Gene 3 (see Figure 4.3 of [46]).

Not all feed-forward loop types appear with equal frequency in regulatory networks (see Figure 4.4 of [46]). The most abundant feed-forward loop is the *type-1 coherent feed-forward loop* (C1-FFL), in which all three links are by activators, and appearing with a frequency of more than 45% among all feed-forward loops in the regulatory network of *Saccharomyces cerevisiae* [46].

In addition to the signs attached to the links, in order to completely determine the dynamics of the feed-forward loop one must also specify how the inputs from the two regulators Gene 1 and Gene 2 are integrated at the promoter of Gene 3, that is, the input function of Gene 3 must be a bi-variate function implementing a *logical gate* at the promoter of Gene 3. There are two biologically reasonable logical gates: (i) the AND logic, in which *both* Gene 1 *and* Gene 2 production need to be high in order to turn ON the expression of Gene 3; (ii) the OR logic, in which *either* Gene 1 *or* Gene 2 production is sufficient to turn ON the expression of Gene 3. The way we implement these logical gates in our formalism is by composing the input function with some appropriate operations at the level of protein concentrations: the operation of sum (+) of the concentrations for the OR logic and the operation of multiplication (×) of the concentrations for the AND logic. It is easy to see that this prescription does indeed satisfies the requirements for the OR logic and the AND logic as explained above.

Thus, there are 8 types of feed-forward loop sign combinations, each of which can appear with at least two types of the logical gate (AND, OR) at the third node, performing a total of 16 distinct coupled RDS associated with the feed-forward loop network topology.

The general system of affine RDS corresponding to a feed-forward loop is the following:

$$\begin{aligned}
u_{n+1} &= (1-\gamma_1)u_n + \delta_1 \zeta_n \\
v_{n+1} &= (1-\alpha_1)v_n + \beta_1 u_n \\
x_{n+1} &= (1-\gamma_2)x_n + \delta_2 \xi[v_n] \\
y_{n+1} &= (1-\alpha_2)y_n + \beta_2 x_n \\
w_{n+1} &= (1-\gamma_3)w_n + \delta_3 \vartheta[v_n, y_n] \\
z_{n+1} &= (1-\alpha_3)z_n + \beta_3 w_n
\end{aligned} \qquad (***)$$

where $(u_n, v_n)$ are the state variables of Gene 1, $(x_n, y_n)$ are the state variables of Gene 2 and $(w_n, z_n)$ are the state variables of Gene 3. The reaction rates of Gene 1 are $\alpha_1, \beta_1, \gamma_1, \delta_1$, the reaction rates of Gene 2 are $\alpha_2, \beta_2, \gamma_2, \delta_2$ and the reaction rates of Gene 3 are $\alpha_3, \beta_3, \gamma_3, \delta_3$. The switching process $\zeta$ of Gene 1 may be set as in a single gene system (IID or Markovian). The links arriving at Gene 2 and Gene 3 are given by two input functions $p_{12}(v) = H(k_1, r_1, v)$ for the coupling between Gene 1 and Gene 2; $q_{(12)3}(v,y) = H(k_2, r_2, v \times y)$ (AND logic) or $q_{(12)3}(v,y) = H(k_2, r_2, v+y)$ (OR logic), for the coupling between Gene 1 and Gene 3 AND/OR the coupling between Gene 2 and Gene 3, respectively. Therefore, $p_{ON} = \mathbf{P}(\xi[v_n] = 1) = p_{12}(v_n)$ (or $p_{12}(\varepsilon_1 - v_n)$) and $q_{ON} = \mathbf{P}(\vartheta[v_n, y_n] = 1) = q_{(12)3}(v_n, y_n)$ (or $q_{(12)3}[(\varepsilon_1 - v_n), (\varepsilon_2 - y_n)]$).

As explained before, in the type-1 coherent feed-forward loop (C1-FFL) with AND logic the production of Gene 3 requires binding of both Gene 1 and Gene 2. This implies that the concentration of Gene 2 must build up to sufficient levels to cross the activation threshold for Gene 3. Thus once the signal of Gene 1 appears, Gene 2 needs to accumulate in order to activate Gene 3. This results in a delay in Gene 3 production. Moreover, there is no delay for the turning OFF of Gene 3, that is, when Gene 1 is turned OFF, Gene 3 production stops at once. This type of behavior makes the C1-FFL with AND logic a "signal-sensitive delay" circuit and is very useful in highly fluctuating environments (like a cell), where stimuli can be present for brief pulses that should not elicit a response. The signal-sensitive delay property of the C1-FFL with AND logic has been extensively discussed by [46] in the deterministic framework, that is, in the small noise limit. Fig. 5 shows the time series of the three genes in a C1-FFL with AND logic and the signature of the signal-sensitive delay property manifests itself in the time series of Gene 3, which displays spikes of production followed by long gaps of absence of activity. In this example, Gene 1 is activated by a slow Markovian switch and Gene 2 is activated by very steep Hill function, making it a very fast switch with bimodal stationary distribution, and showing that the signal-sensitive delay property persists even when one is very far from small noise conditions.

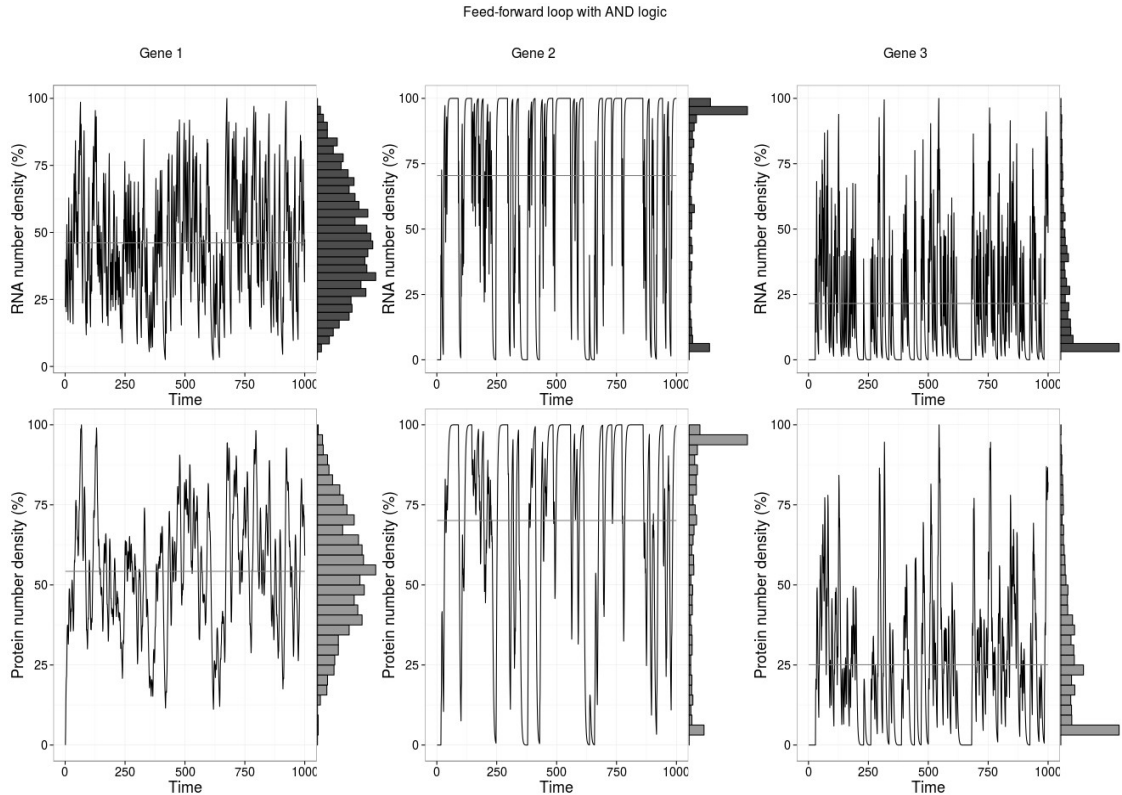

**Figure 5.** Simulation of typical time series and stationary distributions of a feed-forward loop (C1-FFL with AND logical gate). Gene 1 (left) is activated by a slow Markovian switch, Gene 2 (center) is activated by a very steep Hill function, Gene 3 (right) is activated by both Gene 1 AND Gene 2. The time scale (abscissa axis) is of the order of the average time to transcribe 1 gene ($\approx$ ½ min in yeast). The number density scale (ordinate axis) is the relative concentration of molecules (%) with respect to its maximum value. The horizontal line is the mean value of the level of expression.

When the C1-FFL has an OR logical gate at Gene 3 promoter instead of an AND gate the production of Gene 3 is activated immediately upon the activation of Gene 1 because it only takes one input to activate an OR logical gate. Thus there is no delay in Gene 3 following the activation of Gene 1. In contrast, Gene 3 is deactivated with a delay following the deactivation of Gene 1, because both inputs need to go off in order to the OR logical gate be deactivated. That is, the C1-FFL with OR logic is a signal-sensitive delay circuit for turning OFF, whereas the C1-FFL with AND logic is a signal-sensitive delay circuit for turning ON. Hence, the C1-FFL with OR logic can maintain a steady expression of Gene 3 even if the input signal is momentarily lost. The signal-sensitive delay property for turning OFF of the C1-FFL with OR logic has been extensively discussed by [46] in the deterministic framework, that is, in the small noise limit. Fig. 6 shows the time series of the three genes in a C1-FFL with OR logic. In this example, Gene 1, Gene 2 and Gene 3 are exactly as in the example of Fig. 5, except that the logical gate at the promoter of Gene 3 is OR instead of AND. The characteristic feature of the signal-sensitive delay for turning OFF is displayed by the time series of Gene 3 which is quite steady and concentrated near the mean level of expression, even though the expression of Gene 2 is very intermittent, jumping between full activation and full inactivation. Again, this example shows that the signal-sensitive delay property for turning OFF persists even when one is very far from small noise conditions.

Finally, it is worth to remark that both the C1-FFL with AND and OR logic dynamics have been experimentally demonstrated to occur in *E. coli* regulatory network [46] and both are likely to be very frequent in other regulatory networks.

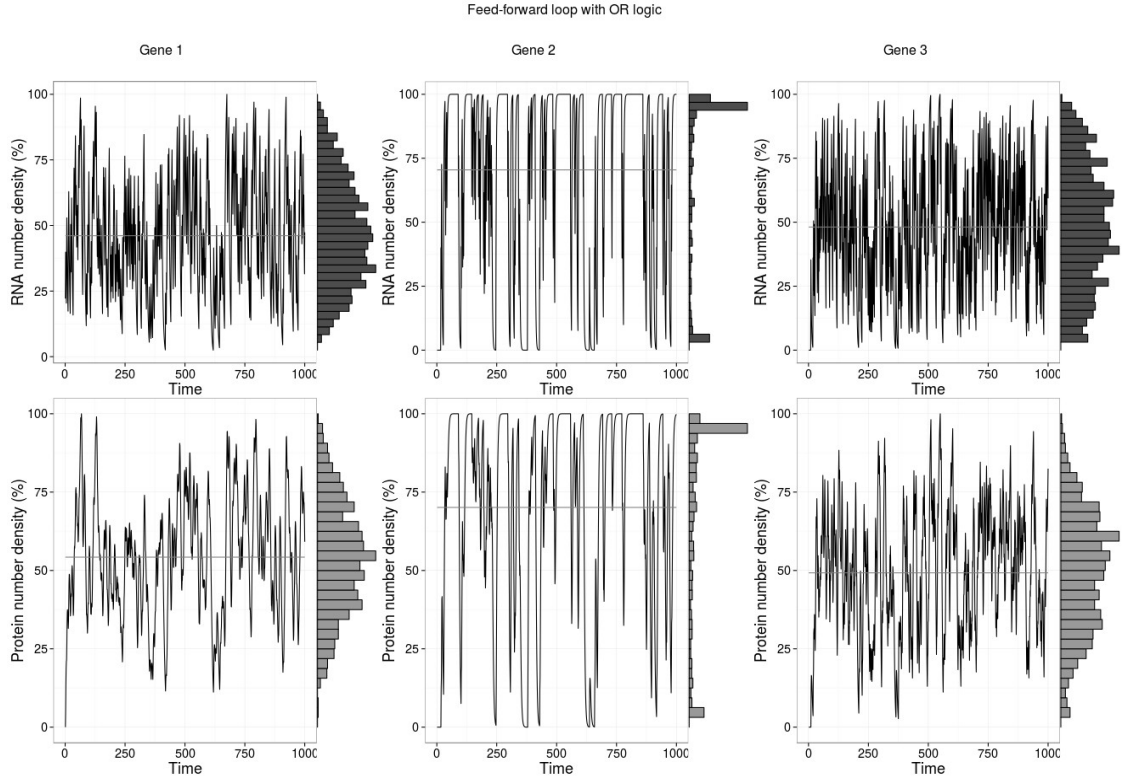

**Figure 6.** Simulation of typical time series and stationary distributions of a feed-forward loop (C1-FFL with OR logical gate). Gene 1 (left) is activated by a slow Markovian switch, Gene 2 (center) is activated by a very steep Hill function, Gene 3 (right) is activated by either Gene 1 OR Gene 2. The time scale (abscissa axis) is of the order of the average time to transcribe 1 gene ($\approx$ ½ min in yeast). The number density scale (ordinate axis) is the relative concentration of molecules (%) with respect to its maximum value. The horizontal line is the mean value of the level of expression.

## 2.4 Computational simulations and fitting of experimental data

The computational simulation and approximation of the time series and the stationary distributions of the discrete time models introduced here can be be efficiently implemented [51]. One of the advantages of our formulation, as concerning the computational aspect, is that there is no need for special methods for its simulation (like the Gillespie's method used for simulation of master equations). Since the equations are discrete in time and the state variables are the observable quantities, all that is necessary is a pseudo-random number generator to compute probabilities and a system to iterate the equations, once the parameters and initial conditions have been chosen. For instance, all the simulations presented here were performed using the software *XPPAUT* [52], following the instructions in [53], and plotted using the *R* software [54] (see Supplementary Material for the codes to generate all the figures).

Regarding the relation of the our formalism with experimental data, the difficulties are the same as the ones faced by the deterministic formalism: the fast growth of the number of parameters that need to be measured experimentally. Each gene in a network needs four parameters (the reaction rates $\alpha$, $\beta$, $\gamma$, $\delta$) and each link needs, at least, two parameters (the Hill function parameters $r$, $k$) – more complex input functions than the ones we used require more than two parameters for each multi-input promoter region. Bearing in mind that experimental confirmation of a link between two nodes in a regulatory network is already a difficult task, it is not surprising that experimentally measured values of the reaction rates and input function parameters with the necessary precision is very scarce.

*3. Conclusions*

Various methods for the analysis of gene regulation have been proposed, depending on the biology of the phenomena. The approach of [55–57] was designed to explore the effects of small number of mRNA and protein molecules in bacteria. In [55] the authors follow the assumptions made in [58,59], that there is a rapidly achieved equilibrium between regulatory proteins and the corresponding gene promoters. The same assumption was made in [56], who applied the stochastic formulation of chemical kinetics developed by [22,23] as a means to analyze the *lambda phage* in *Escherichia coli*.

Stochasticity due to switching of the gene state was first recognized by [60] and then was analyzed by [61]. Their approach involves the Chapman-Kolmogorov equation for probability distribution defined on discrete states, which is then approximated by the Fokker-Planck equation. In the case of a single self-regulated gene, the Fokker-Planck equation is further simplified by neglecting the diffusion term, which leads to a first-order system of PDE's. In [61], the authors also consider a system of two mutual repressors and, assuming that they are identical, compute the marginal distribution of the protein using Monte Carlo simulations. In their recent work, [62] use the Fokker-Planck equation to calculate all first and second moments of the probability density function and focus on reactions leading to the formation of dimers and tetramers. Transcription regulation involving switching between discrete high and low transcription rates was also considered in a frequency domain by [63]. Their approach provides the frequency distribution of noise associated both with mRNA synthesis/degradation and noise resulting from the operator binding events that cause bursts of transcription. Switching between more than two states in the framework of master equations has been considered in [20].

Following [60] and others, our single-gene model focus on stochasticity in eukaryotic gene expression, which occurs at the level of transcription regulation. The approach combines the stochastic switch description of kinetics of reactants present in a small number of copies (in this case gene copies) with discrete-time difference equations for the description for processes involving larger number of reacting molecules (i.e., mRNAs and proteins).

The main advantage of our framework is to overcome a major difficulty of other formalisms (master equation and Langevin) by providing an explicit prescription for building a model for a stochastic coupled regulatory network according a fixed (arbitrary) network topology given the uncoupled dynamics of each node (reaction rate parameters) and the coupling rules between them (Hill coefficients). This advantage becomes critical in the case where there are more that one transcription factor regulating the same gene. For instance, in [49] where the master equation formalism is employed to model some simple network motifs, the authors are able to treat only the motifs with one input link per node. In contrast, we were able to implement and simulate a stochastic version of one of the simplest network motifs with more that one input link per node, to wit, the feed-forward loop, without any extra effort. Finally, the composed stochastic dynamics of a regulatory network reproduces the dynamics of the corresponding deterministic dynamics in the "small noise limit", allowing us to automatically obtain a set of validations (regarding the properties related to the mean level of expression) of the models based on our formalism.

We also have shown that some "truly stochastic" (qualitative and quantitative) phenomena observed in single gene models and simple motif networks, several of them extensively discussed in the literature, are faithfully reproduced by our formalism, thus providing another set of validations of the model that goes beyond average values.

The model proposed here is based on the assumption that the gene promoters, in the time-scale at the order of half-life of the mRNA, are not in a statistical equilibrium. This assumption is supported by a growing number of experiments on single-cell gene expression, showing cell-to-cell heterogeneity in mRNA levels, fluctuations of which are too large to be explained only by the effects of finiteness of number of mRNAs [64,65]. The analytical framework introduced here can be compared with experimental data, since it is possible to tabulate some measurable parameter values from data sets and identify common genetic regulatory elements in genes with similar noise behaviors.

## 4. Methods: The Theoretical Framework

In order to construct dynamical systems on a network one needs two types of structure. The first type is the building block, given by a class of dynamical systems for the nodes of the network, which should have the following properties: (i) a dynamical system in this class can receive input from another dynamical system in the class and (ii) a dynamical system in the class produces an output that is coherent with their "input entry", namely, the output of one dynamical system in the class is meaningful as input to another dynamical system in the class. The second type of structure is given by a prescription of coupling two systems on nodes that are connected by a link and, even more importantly, the specification of how to handle multiple inputs coming from several external sources into the same node.

### 4.1 The dynamics of single gene systems

Let us start by describing an appropriate class of dynamical systems for the nodes. A random version of the *classical rate equation of kinetic theory* is given by a stochastic differential equation (SDE) of the form:

$$\frac{d}{dt}[X] = -K_1[X] + K_2 \xi[Y_{ext}] \qquad (1)$$

where $K_1$, $K_2$ and $Y_{ext}$ are continuous-time stationary stochastic processes. The process $K_1$ is the decay rate, the process $K_2$ is the production rate and $\xi[Y_{ext}]$ is the process of activation-deactivation of transcription, which depends on the expression level $[Y_{ext}]$ of externally produced (upstream) transcription factors acting near the promoter region of the gene.

Choosing an appropriate time step $\Delta t$, which can be interpreted as the time interval between two successive observations of the dynamical variables (recurrence time) and applying Euler's discretization method to equation (1) gives

$$x_{n+1} = (1 - \gamma_n) x_n + \delta_n \xi_n \qquad (2)$$

where $x_n = X(n\Delta t)\Delta t$ is the number density of mRNA molecules produced between two successive observations, $\gamma_n = K_1(n\Delta t)$ is the degradation process between two successive observations, $\delta_n = K_2(n\Delta t)$ is the production process between two successive observations, and are real-valued discrete-time stochastic processes satisfying $0 < \gamma_n, \delta_n < 1$. The switching stochastic process $\xi_n = \xi[Y(n\Delta t)]\Delta t$ determines if the gene is turned ON ($\xi_n$ is on the state 1) or turned OFF ($\xi_n$ is on the state 0). Therefore, $\xi_n$ is a discrete-time finite state stochastic process taking values in {0=OFF, 1=ON}. More generally, $\xi_n$ may be an *m*-ary stochastic process, that is, $\xi_n$ take $m$ possible distinct values in the interval [0,1] equally separated starting from 0 and ending at 1, representing intermediate levels of activation. Furthermore, the only assumption on the stochastic processes considered here is that they are *stationary stochastic processes,* in the sense that their joint probability distributions do not change when shifted in time or space – this is the standard assumption in the theory of random dynamical systems [38].

Hence, we have a discrete-time version of the rate equations with an explicit dependence on the stochastic processes $\gamma_n$, $\delta_n$ and $\xi_n$ without any *a priori* specified type of probability distributions other than the fact that they are stationary. By selecting these stochastic processes based on direct microscopic observation and evidence, one can model specific aspects of the dynamical interaction between the gene (more specifically, its promoter region) with the transcription factors and other regulating enzymes that may act together in a coherent way.

So far, we have defined how a node receives the input; we also have to define how it produces an output that can be used as an input by another node. This can be done by introducing another equation for the protein translated from the mRNA of the gene under consideration. Similar argument leads to the following discrete-time equation

$$y_{n+1} = (1-\alpha_n)y_n + \beta_n x_n \qquad (3)$$

where $y$ is a non-negative real-valued variable representing the protein number density produced between two successive observations, $\beta_n$ is the protein production process and $\alpha_n$ is the protein degradation process satisfying $0 < \alpha_n, \beta_n < 1$. In summary, the class of dynamical systems representing the nodes of our model consists of systems composed by two coupled equations of the form (2) and (3).

Before, proceeding, let us rewrite our basic system of equations in matrix form, which will be useful for constructing networks. By combining the two variables $x$ and $y$ into a vector $(x,y)$ we can write

$$\begin{pmatrix} x \\ y \end{pmatrix}_{n+1} = \begin{pmatrix} 1-\gamma_n & 0 \\ \beta_n & 1-\alpha_n \end{pmatrix} \begin{pmatrix} x \\ y \end{pmatrix}_n + \xi_n \begin{pmatrix} \delta_n \\ 0 \end{pmatrix} \qquad (4)$$

For example, when $\xi_n$ take values on the finite set $\{0, 1/(m-1), 2/(m-1), \ldots, 1\}$ and $\alpha, \beta, \gamma, \delta$ are constant reaction rates, the phase space of system (4) is the rectangle $[0,\delta/\gamma] \times [0,\beta\delta/(\alpha\gamma)]$. More generally, a generic member of our class of dynamical systems can be written as

$$z_{n+1} = A_n z_n + b_n \qquad (5)$$

where $z = (x,y)$, $b_n$ is a random vector and $A_n$ is a two-by-two random matrix. Dynamical systems of the form (5) are called *affine random dynamical systems (affine RDS)*. Since the coefficients $A$ and $b$ are *stationary stochastic processes*, it can be shown that an affine RDS posses a unique *stationary distribution* [38,45], which represents the equilibrium state of the system.

It is important to point out, that even though an affine RDS resembles an *affine deterministic dynamical system*, it can be highly non-linear (in the sense that its equilibrium state resembles a chaotic attractor of a deterministic dynamical system [66]) due to the stochasticity of its coefficients $A$ and $b$.

From now on, for simplicity, we shall assume that the switching process $\xi_n$ is binary taking values on $\{0=\text{OFF}, 1=\text{ON}\}$ and the reaction rates $0 < \alpha, \beta, \gamma, \delta < 1$ are constant. The extension to the more general situation will be addressed in the last part of this section.

*4.2 The coupling between a transcription factor and its promoter region*

Now we will describe how to couple two affine RDS, as described in the previous section by system (4), in such a way that the resulting coupled system still is an affine RDS. In order to simplify matters, we will assume that our affine RDS (4) have constant reaction rates, once we have this particular case the extension to the general situation is not difficult. Given two affine RDS with variables $(u_n, v_n)$ and $(x_n, y_n)$ representing two genes, driven by the stochastic processes $\zeta_n$ and $\xi_n$, respectively. We want to define the protein variable of the first gene as a transcription factor for the second gene. Let us suppose, for the moment, that $\xi_n$ and $\zeta_n$ are binary, taking values in $\{0=\text{OFF}, 1=\text{ON}\}$. Then, in this case, their probability distributions are determined by a single number $0 < p < 1$.

Let $p_1:[0,\varepsilon] \to [0,1]$ be an *input function*, defined on the range $[0,\varepsilon]$ of the protein variable $v$ of the first gene, where $\varepsilon = \beta\delta/(\alpha\gamma)$ is its maximum value, and set $p_0 = 1 - p_1$. Thus, on has that $p_1$ is a non-negative strictly increasing continuous function and $p_0$ is a non-negative strictly decreasing continuous function, the value $p_1(v_n)$ is interpreted as the probability that transcription is turned ON due to the *activation* promoted the protein and the value $p_0(v_n)$ is the probability that transcription is turned OFF.

In order to model the situation where the transcription factor acts by *repression* at the promoter region instead of *activation*, one simply replaces the variable $v_n$ by $(\varepsilon - v_n)$ in the definition of functions $p_1$ and $p_0$: the new function obtained $q_1(v_n) = p_1(\varepsilon - v_n)$, is a non-negative strictly decreasing continuous function and $q_0 = 1 - q_1$ is a non-negative strictly increasing continuous function. Therefore, we shall treat in detail only the activator case in what follows; the repressor case may be easily obtained by a change of variables $v \to v' = (\varepsilon - v)$.

Now we define the probability distributions of the sequence of random variables $\xi_n = \xi[v_n]$ with two states $j=0,1$ by

$$\mathbf{P}(\xi_n[v] = j) = p_j(v_n) \qquad (6)$$

where $v_n$ is the density number of the transcription factor, at time $n$. A set of input functions $\{p_0, p_1\}$ as defined above is generally called set of place dependent probabilities.

**Proposition 4.1.** *Consider a pair of nodes in a network such that there is one directed link connecting them (which may represent an activator or a repressor), let each node be dynamically represented by an affine random dynamical system and let be given a set of input functions. Then equation* (6) *provides a unique way of coupling the associated equations in such way that the resulting system is an affine random dynamical system.*

**Proof.** The affine system (5) together with a set of place dependent probabilities $\{p_0, p_1\}$ satisfying (6) forms an affine iterated function system (IFS), see [51,51]. In [67] it is shown that an affine IFS driven by place dependent probabilities and satisfying some additional conditions [68–70] naturally gives raise to an affine RDS in the sense that of [38]. In fact, the main result of [67] together with the following two observations proves the proposition: (i) a set of input functions $p_j$ defines place-dependent probabilities for the system of affine equations (5); (ii) the rate coefficients must satisfy $0 < \alpha, \beta, \gamma, \delta < 1$, which means that the system of affine equations defines a *contractive mapping* of the interval. Provided the input functions $p_j$ are sufficiently regular, there exists an associated invariant measure to the affine IFS driven by place dependent probabilities [71,72] and the fact that the affine equation (5) is a contractive mapping implies that the associated invariant measure is attractive and, therefore, unique. The existence of a unique invariant measure is sufficient for the main result of [67] to be applied. Finally, it is easy to show that the product system formed by the two coupled single gene systems of the same dimension is again an affine RDS of twice that dimension. □

By iterating the construction presented here one can build affine RDS's associated to some simple network topologies: linear chains; simple cycles, for instance, the repressilator [73,74], which is a simple cycle of repressors; and more generally, any network with input-valency equal to one (only one directed link entering each node).

However, in order to account for the most general network topologies, one needs to specify how to combine two or more inputs entering the same node. Indeed, in order specify a *dynamical network with multiple inputs* it is not sufficient to give the topology of the network, but one also must specify the *logical gating* at the nodes with more that one input – that is, the rules that control how multiple signals are combined into one unique signal.

*4.3 The coupling with multiple transcription factors*

Let us suppose that a node of a gene regulatory network under consideration receives $k$ input arrows, that is, the corresponding gene have $k$ distinct transcription factors $(y_1,...,y_k)$ regulating the corresponding stochastic process $\xi_n$. The random variable $\xi_n$ depends on the number density of the transcription factors through a set of multi-variate input functions $\{p_j\}$, namely, $\mathbf{P}(\xi_n[y_1,...,y_k] = j) = p_j(y_{1,n},...,y_{k,n})$, where $y_{1,n},...,y_{k,n}$ are the number densities of the transcription factors at time $n$. The input functions are defined on the appropriate domain and assume values in the interval $[0,1]$. These multi-variate input functions represent the interactions between the various transcription factors.

There are, at least, two natural *logical gates* for combining the action of two transcription factors into one random variable $\xi_n$ : (i) the OR *logical gate*, denoted by $\xi_n[y_1+y_2]$ and (ii) the AND *logical gate*, denoted by $\xi_n[y_1 \times y_2]$. The *simplest form* of an OR logical gate is given by

$$\mathbf{P}(\xi_n[y_1+y_2] = j) = p_j(y_{1,n}+y_{2,n}) \qquad (7)$$

and the *simplest form* of an AND logical gate is given by

$$\mathbf{P}(\xi_n[y_1 \times y_2] = j) = p_j(y_{1,n} \times y_{2,n}) \qquad (8)$$

Moreover, any combination of the logical operations AND and OR involving more than two variables can be implemented by iterating prescriptions (7) and (8), which gives the minimal number of parameters. Moreover, the transformation $y \to y' = (\varepsilon - y)$ may be seen as an implementation of a NOT *logical gate* which allows to change the plus sign (+) of an activator to the minus (−) of a repressor.

It is possible to define more complicated and biologically motivated functional forms for the multivariate input functions, with several plateaus and thresholds, at the cost of introducing more parameters into the model (see [46]).

The definitions of the logical gates given above are consistent with the standard terminology. For instance, an OR logical gate promotes activation of mRNA production in the presence of any one of the two transcription factors represented by $y_1$ OR $y_2$. On the other hand, an AND logical gate promotes activation of mRNA production only if both transcription factors represented by $y_1$ AND $y_2$ are present, to wit, if one of them is zero then the probability that transcription is turned OFF is equal to one.

**Proposition 4.2.** *Consider a node in the network which has multiple inputs from other k nodes and suppose that corresponding logical gates between each pair of inputs are given by applying one of the two logical operations: AND, OR. Then, given a set of input functions, depending on k variables, the prescriptions given by (7) and (8) provide a unique way of coupling the associated equations in such way that the resulting system is an affine random dynamical system.*

**Proof.** The construction of [67] holds in the case of multi-variate place-dependent probabilities and therefore a similar argument as in the proof of Proposition 4.1 gives the result. □

**Proof of Theorem A.** It follows by iteration of the procedure described in sub-sections 4.2 and 4.3. At each stage Propositions 4.1 and 4.2 guarantee that the coupled system is an affine RDS. The dimension of the full system is given by the total number of state variables which twice the number of nodes. □

Therefore, we have accomplished our first goal, that is, we have shown how one may coherently associate a unique (up assignment of the place-dependent probabilities) affine RDS to any network topology with the specification of the transcription types on all the arrows and the specification of the logical gating on the nodes with multiple inputs. It is important to emphasize that the full system of coupled equations that is built from single gene RDS is also an RDS, a fact that is trivial in the deterministic case but non-trivial at all in the stochastic case. It is the use of place-dependent probabilities that allows one to use the invariant measures of the single gene RDS to construct an invariant measure for the switching process of the coupled RDS corresponding to the full network.

*4.4  Deterministic rate equations as average equations*

In order to obtain the average equations associated to the stochastic model introduced here the first step is the computation of the mean value $\langle \xi \rangle$ and the variance $\sigma^2_\xi$ of the switching process, taking into account that the activation process $\xi_n$ have its probability distribution defined through input functions.

**Proposition 4.3.** *Let $\{\xi_n\}$ be a binary process taking values on $\{0,1\}$ with probability distribution function given by a place-dependent probability $p(y)$, at least twice differentiable, where y denotes the protein variable of an upstream transcription factor. Then the following formulas hold almost surely (in y): $\langle \xi \rangle = p(\langle y \rangle) + O((y - \langle y \rangle)^2)$, $\sigma^2_\xi = p(\langle y \rangle)[1 - p(\langle y \rangle)] + \sigma^2_y [p'(\langle y \rangle)^2 - \langle y \rangle^2 p''(\langle y \rangle)^2] + O((y - \langle y \rangle)^4)$. In particular, the dispersion index of $\xi$ is $D_\xi = 1 - p(\langle y \rangle) + O((y - \langle y \rangle)^2)$ and the coefficient of variation of $\xi$ is $\eta_\xi^2 = (1 - p(\langle y \rangle))/p(\langle y \rangle) + O((y - \langle y \rangle)^2)$.*

**Proof.** When the random variable $\xi_n$ is Bernoulli then $p$ is constant and one has that $\langle \xi_n \rangle = p$ and $\sigma^2(\xi_n) = p(1 - p)$. However, when the probability distribution of $\xi_n$ is given by a place-dependent probability function $p(y)$, one first must compute the expectation of $\xi_n$ conditioned by $y_n$, which is simply $\langle \xi_n | y_n \rangle = p(y_n)$. The (unconditional) expectation $\langle \xi_n \rangle$ of $\xi_n$ is given by the $y$-mean value $\langle \xi_n \rangle = \langle p(y_n) \rangle$. The variance may be obtained in a similar way using the formula for total variance $\sigma^2(\xi_n) = \langle \sigma^2(\xi_n | y_n) \rangle + \sigma^2(\langle \xi_n | y_n \rangle)$, where $\sigma^2(\xi_n | y_n)$ is the conditional variance of $\xi_n$. In our case, this formula gives $\sigma^2(\xi_n) = \langle p(y_n) [1 - p(y_n)] \rangle + \sigma^2(p(y_n))$. Taylor expansion theorem and the assumption that the function $p(y_n)$ is sufficiently continuously differentiable allows one to obtain the following approximations (up to second order), holding almost surely: $\langle p(y_n) \rangle = p(\langle y_n \rangle) + \frac{1}{2} \sigma^2(y_n) p''(\langle y_n \rangle) + O(|y_n - \langle y_n \rangle|^3)$ and $\sigma^2(p(y_n)) = \sigma^2(y_n) (p'(\langle y_n \rangle)^2 - \langle y_n \rangle^2 p''(\langle y_n \rangle)^2) + O(|y_n - \langle y_n \rangle|^4)$. □

**Corollary 4.4.** *Suppose that the reaction rates α, β, γ, δ are constant. Then the classical rate equations of kinetic theory coupled through Hill input functions can be readily obtained from the stochastic model presented here, by taking the expectation values of the state variables in the internal dynamics of each single-gene sub-system of the whole system.*

**Proof of Theorem B.** By taking mean values in equations (2)-(3) and disregarding higher order terms – which, by Corollary 4.4, depend only on higher order moments, which are assumed to be much smaller that the average values – one easily obtains a discrete-time version of the deterministic rate equations for the mean concentrations $[X_n] = \langle x_n \rangle$ and $[Y_n] = \langle y_n \rangle$, and coupled through a deterministic input function denoted as $\langle \xi \rangle [Y_n] = p(\langle y_n \rangle)$. □

This average process can be applied to all the state variables in the full system of equations associated to a gene network giving the correct discrete-time deterministic system (see [46] for more details on deterministic rate equations). Since the higher order terms of the expansions in proposition 4.3 contain all the higher order moments, one may regard the deterministic equations as the "small noise limit" of the original stochastic equations, that is, when the second moment (and consequently the higher moments) are much smaller that the average values the stochastic evolution is dominated by the average dynamics which is exactly the dynamics given by the deterministic rate equations.

*4.5 Statistical properties of single-gene systems*

Let us now turn to the analysis of single-gene system and study the statistical properties of the state variables $x_n$ and $y_n$. Recall that the process $\xi_n$ is stationary and therefore its moments become time-independent when the initial condition $\xi_0$ is drawn from the equilibrium distribution. In particular its mean value $\mu_\xi = \langle \xi \rangle$ and variance $\sigma^2_\xi = \sigma^2(\xi) = \langle \xi^2 \rangle - \langle \xi \rangle^2$ are time-independent. Moreover, since our equations are affine, it follows that the processes $x_n$ and $y_n$ have unique equilibrium distributions [38,45]. Hence, they are stationary processes and their moments become time-independent, as long as the initial conditions are identical to the equilibrium distributions.

Defining $\mu_x = \langle x \rangle$ as the mean value of $x$ at equilibrium and taking expectation values in both sides of equation (2) gives

$$\mu_x = \frac{\delta}{\gamma} \mu_\xi \qquad (9)$$

It is worthwhile to remark that, in the case where the process $\xi_n$ is Bernoulli, formula (9) coincides with the expected level of transcription in the continuous-time version of the model obtained in [40] (equation (2)) which agrees with the classical result, as well. The translation of notation between our equation (9) and theirs is the following: $\langle \xi \rangle = nP$, $\gamma = h$ and $\delta = a$.

Turning to the protein equation (3) and defining $\mu_y = \langle y \rangle$ as the mean value of $y$ at equilibrium, the same reasoning as before gives

$$\mu_y = \frac{\beta}{\alpha} \mu_x = \frac{\delta \beta}{\gamma \alpha} \mu_\xi \qquad (10)$$

Now, one may compute the variance $\sigma^2_x$ of $x$ at equilibrium by squaring equation (2) and taking mean values on both sides:

$$\sigma^2_x = \frac{\delta^2 \sigma^2_\xi + 2\delta(1-\gamma)\sigma_{\xi x}}{1-(1-\gamma)^2} \qquad (11)$$

where $\sigma_{\xi x} = \langle \xi x \rangle - \langle \xi \rangle \langle x \rangle$ is the *cross-covariance between $\xi$ and $x$* at the equilibrium. Similarly, the variance $\sigma^2_y$ of $y$ at equilibrium is given by

$$\sigma^2_y = \frac{\beta^2 \sigma^2_x + 2\beta(1-\alpha)\sigma_{xy}}{1-(1-\alpha)^2} \qquad (12)$$

where $\sigma_{xy} = \langle xy \rangle - \langle x \rangle \langle y \rangle$ is the *cross-covariance between $x$ and $y$* at the equilibrium.

It is not difficult to show that the general time dependent cross-correlation functions $\langle \xi_n x_n \rangle$ and $\langle x_n y_n \rangle$ can be written terms of the *temporal auto-correlation functions* $\langle \xi_n \xi_{(n-1)-i} \rangle$ multiplied by powers of $(1-\gamma)$ and $(1-\alpha)$. In fact, it follows from the following general formulas

$$x_n = (1-\gamma)^n x_0 + \delta \sum_{i=0}^{n-1} \xi_{(n-1)-i}(1-\gamma)^i \qquad (13)$$

for the evolution of $x_n$ given a initial condition $x_0$ and

$$y_n = (1-\alpha)^n y_0 + \beta \sum_{i=0}^{n-1} x_{(n-1)-i}(1-\alpha)^i \qquad (14)$$

for the evolution of $y_n$ given a initial condition $y_0$.

Since the process $\xi_n$ is stationary, the temporal auto-correlation functions $\rho_\xi(n) = \langle \xi_k \xi_{k+n} \rangle$ actually depend only on the time lags, namely, the absolute value of the differences between the instants of time. From equation (13) and the stationarity of the process $\xi_n$, one concludes that the temporal auto-correlation function satisfies $\rho_\xi(n) = \langle \xi_k \xi_{k+n} \rangle = \langle \xi_k \xi_{k-n} \rangle = \rho_\xi(-n)$ and so one finds (assuming that the initial conditions $x_0$, $y_0$ and the variables $\xi_n$ are independent for all $n > 0$)

$$\langle \xi x \rangle = \delta \sum_{n=0}^{\infty} \rho_\xi(n+1)(1-\gamma)^n \qquad (15)$$

Similarly, $\langle xy \rangle$ is given, in terms of auto-correlation functions $\rho_x(n) = \langle x_k x_{k+n} \rangle$, by

$$\langle xy \rangle = \beta \sum_{n=0}^{\infty} \rho_x(n+1)(1-\alpha)^n \qquad (16)$$

As mentioned before, affine RDS like the ones given by equations (4) or (5) possess a unique *stationary distribution* or *invariant probability measure*, which carry information about of the asymptotic behavior of the dynamics – actually, the mean and variance computed above are simply the mean and variance of the aforesaid stationary distribution. Although it is often impossible to find explicit formulas for the density of the stationarity distributions – in fact, they usually have fractal structure and in several cases are singular with respect to the Lebesgue measure – it is very easy to compute approximations to the them as we have done in Section 2.

*4.6    The single-gene with IID switching*

The simplest switching mechanism that may be incorporated in our formalism is the *IID switching*, to wit, the process $\xi_n$ is a *binary Bernoulli process* with (constant) probability $p$ of turning ON and $(1 − p)$ of turning OFF ($0 < p < 1$). In this case, the process $\xi_n$ is simply a sequence of independent, identically distributed (IID) random variables taking values on {0=OFF, 1=ON} with probability distribution function $f(j) = p^j (1 − p)^{j−1}$, for $j = 0,1$. In this case, one can compute the first and second moments of the three state variables, taking into account that $\rho_\xi(n)=0$ for all $n$.

**Proposition 4.5.** *Suppose that the switching process $\xi_n$ is a binary Bernoulli process with probability $p$ and the reaction rates α, β, γ, δ are constant. Suppose that the initial conditions $x_0$ and $y_0$ have independent distributions and both are independent from $\xi_n$. Then the 2-dimensional process $z = (x,y)$ is Markovian, the temporal auto-correlation function $\rho_x(n)$ is given by $\rho_x(n) = \langle \xi^2 \rangle \delta^2 / (1−(1−\gamma)^2)(1−\gamma)^{|n|}$ and $\rho_y(n)$ is given by $\rho_y(n) = \langle x^2 \rangle \beta^2/(1 − (1 − \alpha)^2) (1 − \alpha)^{|n|}$. The cross-correlation function is*

$$\langle x\,y \rangle = \frac{\beta \delta^2 (1-\gamma) \langle \xi^2 \rangle}{(1-(1-\gamma)^2)(1-(1-\gamma)(1-\alpha))} \qquad (17)$$

**Proof.** In general, any random dynamical system driven by an IID stochastic process is always Markovian (see [37,38]). Moreover, the 2-dimensional stochastic process $z = (x,y)$ is Markovian. The auto-correlation functions are easily obtained from eqs. (13) and (14) and the cross-correlation from eq. (16). □

**Proposition 4.6.** *Suppose that the switching process $\xi_n$ is a binary Bernoulli process with probability $p$ and the reaction rates α, β, γ, δ are constant. Suppose that the initial conditions $x_0$ and $y_0$ have independent distributions and both are independent from $\xi_n$. Then, at the equilibrium, we have the following:*
(a) $\mu_\xi = p$
(b) $\sigma^2_\xi = p(1 − p)$
(c) $\mu_x = \delta/\gamma\, \mu_\xi$
(d) $\mu_y = \varepsilon\, \mu_\xi = (\beta\delta)/(\alpha\gamma)\, \mu_\xi$
(e) $\sigma_{\xi,x} = 0$
(f) $\sigma^2_x = \delta^2/(1−(1−\gamma)^2)\, \sigma^2_\xi$
(g) $\sigma_{x,y} = \delta^2/(1−(1−\gamma)^2)\, \beta(1−\gamma)/(1−(1−\gamma)(1−\alpha))\, \sigma^2_\xi$
(h) $\sigma^2_y = \delta^2\beta^2/[(1−(1 − \gamma)^2)(1 − (1 − \alpha)^2)]\, [(1+(1−\gamma)(1−\alpha))/(1−(1−\gamma)(1−\alpha))]\, \sigma^2_\xi$

**Proof.** Items (c) and (d) follow from equations (9) and(10), respectively. Item (e) follows from form equation (15) and the fact that $\rho_\xi(|n|)=0$ for all $n$. Item (e) follows from equation (12). Item (g) follows from equation (16) and item (e). Item (h) follows from equation (12) and item (g). □

**Corollary 4.7.** *Suppose that $\xi_n$ is a binary Bernoulli process and the reaction rates α, β, γ, δ are constant. Suppose that the initial conditions $x_0$ and $y_0$ have independent distributions and both are independent from $\xi_n$. Then, at the equilibrium, we have the following:*
(a) $D_\xi = (1−p)$
(b) $D_x = D_\xi\, \gamma\delta/(1−(1−\gamma)^2)$
(c) $D_y = D_\xi\, \alpha\beta\gamma\delta/[(1−(1−\gamma)^2)(1−(1−\alpha)^2)]\, [(1+(1−\gamma)(1−\alpha)) / (1−(1−\gamma)(1−\alpha))]$
(e) $\eta_\xi^2 = (1−p)/p$
(f) $\eta_x^2 = \eta_\xi^2\, \gamma^2/(1−(1−\gamma)^2) \approx \gamma\, \eta_\xi^2$
(g) $\eta_y^2 = \eta_\xi^2\, (\gamma\alpha)^2/[(1−(1−\gamma)^2)(1−(1−\alpha)^2)][(1+(1−\gamma)(1−\alpha))/(1−(1−\gamma)(1−\alpha))] \approx \gamma\alpha\, \eta_\xi^2$

**Proof.** Follows directly from proposition 4.6. In the last part, we have used the secant approximation $x^2/(1−(1−x)^2) \approx x$, since we have that $0 \leq x^2/(1−(1−x)^2) \leq x \leq 1$ and $|x−x^2/(1−(1−x)^2)| \leq 2−\sqrt{2} \approx 0.58$ for all $0 \leq x \leq 1$. □

*4.7   The gene with Markovian switching*

An immediate extension of the gene with IID switching is the gene with Markovian switching, that is, $\xi_n$ is a two-state Markov chain taking values in {0=OFF, 1=ON}. The Markov process $\xi_n$ is completely determined by its transition matrix $P$, a two-by-two stochastic matrix

$$P = \begin{pmatrix} 1-p & p \\ q & 1-q \end{pmatrix} \qquad (20)$$

where $0 < p,q < 1$ and $q \neq 1-p$ (the equality gives the IID case). Here, $p$ is the probability of transition from state 0 (OFF) to state 1 (ON) and $q$ is the probability of transition from state 1 (ON) to state 0 (OFF). The row vector given by $\pi = (\pi_0 = q/(p+q), \pi_1 = p/(p+q))$ is the equilibrium distribution of $\xi_n$, that is, $\pi P = \pi$. Therefore, at equilibrium the expectation value of $\xi_n$ is $\mu_\xi = \pi_1 = p/(p+q)$, the variance is $\sigma^2_\xi = \pi_1\pi_0 = \pi_1(1-\pi_1) = pq/(p+q)^2$, the dispersion index is $D_\xi = \pi_0 = q/(p+q)$ and the coefficient of variation is $\eta_\xi^2 = (1-\pi_1)/\pi_1 = q/p$. Finally, the Markov property of the process $\xi_n$ implies that the auto-correlation function is $\rho_\xi(n) = (1-p-q)^{|n|}$ for all $n \neq 0$ and $\rho_\xi(0) = \sigma^2_\xi = pq/(p+q)^2$.

The mean values of the mRNA variable $x$ and protein variable $y$ at equilibrium are given by inserting the value $\mu_\xi = \pi_1 = p/(p+q)$ into equations (9) and (10).

**Proposition 4.8.** *Suppose that the switching process $\xi_n$ is stationary and the reaction rates $\alpha$, $\beta$, $\gamma$, $\delta$ are constant. Then, the auto-correlation function $\rho_x(n)$ is*

$$\rho_x(n) = \delta^2 \sum_{k=0}^{\infty} \left( \sum_{j=0}^{k} \rho_\xi(n-k+2j) \right)(1-\gamma)^k \qquad (18)$$

**Proof.** This general formula follows from equation (13). By multiplying the equation for $x_k$ by itself $x_{n+k}$ with a shift of $n$ units of time, taking the average of both sides, taking the limit $k \to \infty$ on both sides and rearranging the terms. □

**Corollary 4.9.** *Suppose that the switching process $\xi_n$ is a binary Markov chain determined by a two-by-two stochastic matrix P given by equation* (18) *and the reaction rates $\alpha$, $\beta$, $\gamma$, $\delta$ are constant. Suppose that the initial conditions $x_0$ and $y_0$ have independent distributions and both are independent from $\xi_n$. Then at equilibrium, up to first order in temporal auto-correlation, one has the following.*
(a) *The variance $\sigma^2_x$ of $x$ is $\sigma^2_x \approx \delta^2/(1-(1-\gamma)^2)\,[\sigma^2_\xi + 2(1-\gamma)\rho_\xi(1)]$ and the variance $\sigma^2_y$ of $y$ is $\sigma^2_y \approx \delta^2\beta^2/[(1-(1-\gamma)^2)(1-(1-\alpha)^2)]\,[\sigma^2_\xi + 2((1-\gamma)+(1-\alpha)(1-(1-\gamma)^2))\rho_\xi(1)]$.*
(b) *The dispersion indexes of mRNA and protein are, respectively, $D_x \approx \gamma\delta/(1-(1-\gamma)^2)\,[D_\xi + 2(1-\gamma)\rho_\xi(1)/\mu_\xi]$ and $D_y \approx \alpha\beta\gamma\delta/[(1-(1-\gamma)^2)(1-(1-\alpha)^2)]\,[D_\xi + 2((1-\gamma)+(1-\alpha)(1-(1-\gamma)^2))\rho_\xi(1)/\mu_\xi]$.*
(c) *The coefficients of variation of mRNA and protein are, respectively, $\eta_x^2 \approx \gamma\,[\eta_\xi^2 + 2(1-\gamma)\rho_\xi(1)/\mu_\xi^2]$ and $\eta_y^2 \approx \gamma\alpha\,[\eta_\xi^2 + 2((1-\gamma)+(1-\alpha)(1-(1-\gamma)^2))\rho_\xi(1)/\mu_\xi^2]$.*

*4.8   The auto-regulated motif or self-regulated gene*

Consider the situation where a gene is such that the probability distribution of the switching process $\xi_n$ explicitly depends on the number density $y_n$ of the protein molecules produced by the gene itself. This coupling can be described in our framework by employing an input function $p:[0,\varepsilon] \to [0,1]$ to define the probability distribution of the binary process $\xi[y_n]$ as $\mathbf{P}(\xi_n[y_n] = j) = p_j(y_n)$ $(j = 0,1)$, that is, $p_1 = p(y)$, $p_0 = 1 - p(y)$. From proposition 2.3 and proposition 2.5 we obtain that the mean values of the state variables are: $\langle\xi\rangle \approx p(\langle y\rangle)$, $\langle x\rangle \approx \delta/\gamma\,p(\langle y\rangle)$ and $\langle y\rangle \approx \varepsilon\,p(\langle y\rangle)$, where $\varepsilon = (\beta\delta)/(\alpha\gamma)$. Therefore, the mean value $\langle y\rangle$ is approximately given as the solution of the *fixed-point equation* $y = h(y)$, where $h(y) = \varepsilon\,p(y)$ is continuous function mapping the interval $[0,\varepsilon]$ into itself.

By *Brower's fixed-point theorem* there is always at least one fixed point of $y = \varepsilon\,p(y)$ on $[0,\varepsilon]$ and if we denote the appropriate fixed point by $\bar{y}$ then $\langle y\rangle \approx \bar{y}$, $\langle x\rangle \approx \alpha/\beta\,\bar{y}$ and $\langle\xi\rangle \approx \bar{y}/\varepsilon$. There are two distinct cases that should be taken into consideration: (i) $h$ is strictly decreasing on $[0,\varepsilon]$ and (ii) $h$ is strictly increasing on $[0,\varepsilon]$ – note that we do not suppose that $h$ is surjective.

**Proposition 4.10.** *Consider a continuous monotonic injective function h mapping the interval $[0,\varepsilon]$ into itself. If h is strictly decreasing then there is a unique solution $\bar{y}$ of the fixed-point equation $y = h(y)$ on $[0,\varepsilon]$ such that $0 < \bar{y} < \varepsilon$. If h is strictly increasing then the fixed-point equation $y = h(y)$ may have at most 3 solutions on $[0,\varepsilon]$ and the following configurations are possible:*

(a) *If there is 1 fixed point $\bar{y}$ then the graph of h is entirely below or entirely above the diagonal line, except at the fixed point $y=0$ or $y=\varepsilon$, respectively; or the graph of h crosses the diagonal at exactly the unique fixed point $\bar{y}$.*

(b) *If there are 2 distinct fixed points then the graph of h is entirely below or entirely above the diagonal line, except at the attracting fixed point $y=0$ or $y=\varepsilon$, respectively, and is tangent to the graph of h at the neutral fixed point $\bar{y}_N$; or the graph of h crosses the diagonal line at the attracting fixed point $\bar{y}_A$ and the diagonal line is tangent to the graph of h at the neutral fixed point $\bar{y}_N$.*

(c) *If there are 3 distinct fixed points $0 \leq \bar{y}_{\inf} < \bar{y}_{\mathrm{mid}} < \bar{y}_{\sup} \leq \varepsilon$ then both $\bar{y}_{\inf}$ and $\bar{y}_{\sup}$ are attracting and $\bar{y}_{\mathrm{mid}}$ is a repelling.*

**Proof.** The proposition follows easily from the analysis of the possible configurations between the graph of a continuous monotonic injective function and the diagonal line on the square $[0,\varepsilon]^2$, both when the function is increasing and decreasing.  □

    We also may approximate the corresponding variances $\sigma^2_\xi$, $\sigma^2_x$ and $\sigma^2_y$ using the formulas in propositions 4.6. If we let $\langle y \rangle \approx \bar{y}$ and $F$ be the pre-factor from formula 4.6(h) then $\sigma^2_y \approx p(\bar{y})(1 - p(\bar{y}))$ $[F/(1 - F(p'(\bar{y})^2 - \bar{y}^2 p''(\bar{y})^2))]$. Now, since $\bar{y}$ is a fixed point of $h$, it follows that $p(\bar{y}) = \bar{y}/\varepsilon$ and thus $p(\bar{y})(1 - p(\bar{y})) = \bar{y}/\varepsilon (1 - \bar{y}/\varepsilon)$. The term $(p'(\bar{y})^2 + \bar{y}^2 p''(\bar{y})^2)$ increases when the steepness of the function $p$ increases, due to the fact that $\bar{y}$ gets closer to the inflexion point of $p$ (the point of maximum of the derivative $p'$) making both $p'$ and $p''$ become very large at $\bar{y}$.

**Corollary 4.11.** *Consider an auto-regulated motif coupled to itself through an input function $p(y)$. If the self-coupling acts as a repressor then there is a unique fixed point $\bar{y}$ of the input function $p$ on the interval $[0,\varepsilon]$ such that $\langle y \rangle \approx \bar{y}$. If the self-coupling acts as an activator then the the input function $p(y)$ may have at most 3 fixed points on $[0,\varepsilon]$ and according to the number of fixed points we have the following possibilities: (a) if there is 1 fixed point $\bar{y}$ then $\bar{y}=0$ or $\bar{y}=\varepsilon$ or $0 < \bar{y} < \varepsilon$ and $\langle y \rangle \approx \bar{y}$; (b) if there are 2 fixed points then one of them is attracting $\bar{y}_A$ and the other is neutral $\bar{y}_{RN}$ with $0 \leq \bar{y}_A < \bar{y}_N < \varepsilon$ or $0 < \bar{y}_N < \bar{y}_A \leq \varepsilon$ and $\langle y \rangle \approx \bar{y}_A$; (b) if there are 3 fixed points then $0 \leq \bar{y}_{\inf} < \bar{y}_{\mathrm{mid}} < \bar{y}_{\sup} \leq \varepsilon$ with both $\bar{y}_{\inf}$ and $\bar{y}_{\sup}$ attracting and $\bar{y}_{\mathrm{mid}}$ repelling, with $\langle y \rangle \approx \frac{1}{2}(\bar{y}_{\inf}+\bar{y}_{\sup})$ and the stationarity distribution of y is the mixture of two unimodal distributions with modes near $\bar{y}_{\inf}$ and $\bar{y}_{\sup}$, respectively, reflecting the bistability of the system in this case. Moreover, $\sigma^2_y \approx \bar{y}/\varepsilon(1-\bar{y}/\varepsilon)[F/(1 - F(p'(\bar{y})^2 + \bar{y}^2 p''(\bar{y})^2))]$, $\sigma^2_\xi \approx 1/F \sigma^2_y$ and $\sigma^2_x \approx E/F \sigma^2_y$, where $\bar{y}$ is a fixed point of h. Here F is the pre-factor from formula 4.6(h) and E is the pre-factor from formula 4.6(f).*

    Let us compute the mean values and variances for the classical family of Hill functions given by $p(y)=H(r,k,y)=y^r/(k + y^r)$, where $k$ is called the *dissociation constant* and $r$ is the *Hill coefficient* (see Section 2.1). The fixed-point equation $y = h(y)$ then becomes, in the repressor case, $(\varepsilon - y)^{r+1} - ky = 0$ and, in the activator case, $y^r(\varepsilon - y) - ky = 0$. Observe that one must have $\varepsilon > 1$ in order to ensure existence of a fixed point.

    Both equations can be exactly solved for the Michaelis-Menten kinetics ($r = 1$). In the activator case, there is exactly one solution in the interval $[0,\varepsilon]$ given by $y = \varepsilon + k/2 (1 - \sqrt{1 + 4\varepsilon/k})$. In the repressor case, there are at most two solutions in the interval $[0,\varepsilon]$ given by $y = 0$ and $y = \varepsilon - k$ and, in particular, the non-zero solution exists only if $0 < k < \varepsilon$. Note that there are only two solutions because the $r = 1$ case is not sigmoidal. In fact, the Hill function is sigmoidal only when $r > 1$, when $r \leq 1$ it is strictly increasing. For example, when $r = 2$ the Hill function is sigmoidal and, in the activator case, the equation can be solved explicitly, as well, giving three distinct solutions: the trivial solution $y = 0$ and the double solution $y = \varepsilon/2 (1\pm\sqrt{1 - 4k/\varepsilon^2})$, except when $\varepsilon = 2\sqrt{k}$, in which case the solutions become $y = 0$ and $y = \sqrt{k}$.

In general, for the repressor case, assuming that $\varepsilon > 1$, there is only one solution and, when $r \to +\infty$, it is easy to see, from the fact that $y^{1/r} \to 1$ for all $y > 0$, that $\bar{y} \approx \varepsilon - 1$ ($\bar{y} < \varepsilon - 1$) and the probability distribution of the protein variable is unimodal. On the other hand, for the activator case, $y_{\inf} = 0$ is always a solution and, when $r \to +\infty$, it is easy to see that there are two other solutions in the interval $[0,\varepsilon]$, the largest solution is given by $\bar{y}_{\sup} \approx \varepsilon$, the smaller solution is given by $\bar{y}_{\mathrm{mid}} \approx 1$ ($\bar{y}_{\mathrm{mid}} > 1$). Therefore, one has that $\langle y \rangle \approx \varepsilon - 1$ in the repressor case and $\langle y \rangle \approx \varepsilon/2$ in the activator case.

As for the variances, since $\langle y \rangle \approx \varepsilon - 1$ then $\langle y \rangle/\varepsilon \, (1-\langle y \rangle/\varepsilon) \approx 1/\varepsilon \,(1-1/\varepsilon)$ in the repressor case and since $\langle y \rangle \approx \frac{1}{2}$ then $\langle y \rangle/\varepsilon \,(1-\langle y \rangle/\varepsilon) \approx \frac{1}{4}$ activator case. The form of the factor $[F/(1 - F\,(p'(\bar{y})^2 - \bar{y}^2 p''(\bar{y})^2))]$ depends on which case one is considering: it is of the form $O(1/(\varepsilon^3 r^4))$ in the repressor case and of the form $O(\varepsilon^4/r^4)$ in the activator case. The inflexion point of the Hill function in the repressor case it is $\varepsilon - y_* \approx \varepsilon - 1$ and in the activator case is $y_* = (k(r-1)/(r+1))^{1/r} \approx (k)^{1/r} \approx 1$. Therefore, one has that $\sigma^2_y \approx 1/\varepsilon^4 \,(1-1/\varepsilon)\, O(1/r^4)$ in the repressor case and $\sigma^2_y \approx \frac{1}{4}\, \varepsilon^3\, O(1/r^4)$ in the activator case. Finally, the dispersion indexes of the protein are $D_y \approx 1/\varepsilon^5\, O(1/r^4)$ in the repressor case and $D_y \approx \frac{1}{2}\, \varepsilon^2\, O(1/r^4)$ in the activator case.

**Corollary 4.12.** Consider *a self-regulated gene coupled to itself through a classical Hill function $p(y) = H(r,k,y)$ (in particular, $\varepsilon > 1$). In the repressor case, one has that $\langle y \rangle \approx \varepsilon - 1$, $\langle x \rangle \approx \delta/\gamma - \alpha/\beta$, $\langle \xi \rangle \approx 1 - 1/\varepsilon$ and the probability distribution of the protein is unimodal. In the activator case with three solutions, one has that $\langle y \rangle \approx \varepsilon/2$, $\langle x \rangle \approx \varepsilon\, \alpha/(2\beta)$, $\langle \xi \rangle \approx \frac{1}{2}$ and the probability distribution of the protein is bimodal with one mode near 0 and the other mode near $\varepsilon$. The corresponding dispersion indexes are given by $D_\xi \approx 1/(F\varepsilon^4)\, O(1/r^4)$, $D_x \approx E/(F(\delta/\gamma - \alpha/\beta))(\varepsilon-1)/\varepsilon^4\, O(1/r^4)$, $D_y \approx 1/\varepsilon^5\, O(1/r^4)$ in the repressor case and $D_\xi \approx \frac{1}{2}\, \varepsilon^3/F\, O(1/r^4)$, $D_x \approx (E\beta)/(F\alpha)\, \varepsilon^2$, $D_y \approx \frac{1}{2}\, \varepsilon^2\, O(1/r^4)$ in the activator case. The corresponding coefficients of variation are given by $\eta_\xi^2 \approx 1/(F\varepsilon^3(\varepsilon-1))\, O(1/r^4)$, $\eta_x^2 \approx E/(F(\delta/\gamma - \alpha/\beta)^2)\, (\varepsilon-1)/\varepsilon^4\, O(1/r^4)$, $\eta_y^2 \approx 1 - 1/\varepsilon^4\, O(1/r^4)$ in the repressor case and $\eta_\xi^2 \approx \varepsilon^3/F\, O(1/r^4)$, $\eta_x^2 \approx (E\beta^2)/(F\alpha^2)\, \varepsilon$, $\eta_y^2 \approx \varepsilon^2\, O(1/r^4)$ in the activator case. Here F is the pre-factor from formula 4.6(h) and E is the pre-factor from formula 4.6(f).*

*4.9  General remarks about the mathematical formalism*

In this last Section we briefly expand the discussion regarding the main mathematical aspects of the models introduced here and indicate possible natural extensions and generalizations of the formalism.

**Motivation for the RDS formulation of gene expression dynamics.** In [36,75] the expression of individual genes were analyzed using a linear stochastic differential equation (SDE) inspired by the mass action dynamics (rate equation kinetics). The authors of [36] consider the variation of the expression level of mRNA of a particular gene in different cells in a population under identical growth conditions. Assuming that there is a mean value and a variance typical of each gene in a given population and that the expression level is randomly distributed in the population, let $[X]$ be the dynamic variable representing the expression level of mRNA of a typical gene in the whole population. The mean expression level may increase as a result of the set of chemical reactions leading to the synthesis of mRNA from one side, and may decrease due to mRNA degradation from another side. The time variation of the mean expression level (and its variance) is an intrinsic property of each gene and, under the assumptions expounded above, the model is given by a *linear stochastic differential equation* (SDE), see [36] for details. The linear SDE takes into consideration all causes that affect the expression level of the gene. Among these causes, one can distinguish (at least) two major ones: (i) degradation (decay) and (ii) activation by transcription factors. If one considers only these two effects then one obtains a much simpler equation, namely, a *random version* of the *classical rate equation of kinetic theory*, given by equation (1). A random differential equation for gene expression similar to equation (1) has been proposed in [40,41], although with a slightly different interpretation (see also [76]). The dependence of the switching process in terms of the concentration of the upstream transcription factor is defined through an *input function*, namely, a strictly monotonic continuous function defined on the range of the concentration of upstream transcription factor, which is some bounded interval of real numbers.

**Affine Random Dynamical Systems.** The discrete-time rate equations obtained here are examples of discrete-time random dynamical system (RDS) (see [37,38,77–79]). In that way, several fundamental properties of the model (like absolute continuity or singularity of invariant densities, characteristic exponents, decay rates, auto-correlations and ergodicity) can be addressed and associated in a consistent way to specific properties of the gene in relation to its function. In this framework the modeling of gene expression dynamics is concentrated on identifying / designing a set of appropriate rules and properties of stochastic processes that fit the experimental facts and evidences. As noted before, an affine RDS resembles a *linear deterministic dynamical system*, however it has stochastic coefficients **A** and **b**, which are required to be *stationary stochastic processes*, in the sense that their joint probability distributions do not change when shifted in time or space. An important consequence of stationarity is that parameters such as *mean value* and *variance*, if they exist, also do not change over time or position. In particular, any time-homogeneous Markov process with an equilibrium distribution and arbitrarily long memory is a stationary process.

**Affine RDS, IFS and Bernoulli convolutions.** In the simplest cases where the matrix $A_n$ is constant and $b_n$ is a *Bernoulli processes* (IID sequences of binary random variables), the affine RDS given by equation (5) reduces to the well known *affine iterated function systems* (affine IFS), which have been extensively studied in the last thirty years, see [80,81]. In fact, the *self-similar measures* associated to one-dimensional affine IFS's have been investigated since the 1940's [82,83] until today [84] (see [85] for a comprehensive survey). From this point of view the stationary distributions of $x$ (mRNA) and $y$ (protein) may be characterized in the following way. Starting with equations (13) and (14), assuming that $\{\xi_n\}$ is identically distributed for all $n$ (this is a consequence of the stationarity of the process $\{\xi_n\}$) and taking the limit as $n \to \infty$ on both sides, leads to explicit expressions for the random variables $x$ and $y$ as *exponential moving averages*:

$$x = \delta \sum_{n=0}^{\infty} \xi_n (1-\gamma)^n$$

and

$$y = \beta \delta \sum_{n=0}^{\infty} \xi_n \left( \sum_{k=0}^{n} (1-\gamma)^n (1-\alpha)^{n-k} \right)$$

Observe that, in the case of the variable $x$, the stationary distribution essentially will depend on the parameter $\gamma$ (in the case of the variable $y$ it will depend on the two parameters $\gamma$ and $\alpha$) and on the probability distribution of the stationary variable $\xi_n$, which is the same for all $n$. Another important remark is that the stationary distributions obtained above have compact support in the interval $[0,\varepsilon]$, as opposed to the stationary distributions obtained with the master equations approach, which display an exponential tail reminiscent from the "Poissonain" character of the formalism. As regarding the "form" of the corresponding probability distributions, there is a rich array of possibilities. For example, when $\xi_n$ is a binary IID process taking values on $\{0,1\}$ with probability $p$ for the state 1 and $(1-p)$ for the state 0, if one has $(1-\gamma) < p^p(1-p)^{(1-p)}$ then the distribution of $x$ is concentrated on a Cantor set of Lebesgue measure zero (hence the probability distribution of $x$ does not even have a density with respect to the Lebesgue measure, and it has no atom either). Furthermore, this follows only from the fact that each random variable $\xi_n$ assumes the values 0 and 1, and not on other properties of the process $\xi_n$. An interesting case occurs when $\xi_n$ is a Bernoulli process with $p = 1/2$ and $\gamma = 1/2$. Then one recognizes the usual binary expansion of a random number between 0 and 1 and hence $x$ is uniformly distributed on the interval $[0,1]$. More generally, what we have just described very informally is often called a *Bernoulli convolution* and, even in the IID case, the limiting object is quite complicated and interesting since it involves some deep number theoretic properties [81,85].

**Place dependent probability distributions.** The extension of the formalism to case with more than two states, where $\xi_{w,n}[y]$ is $m$-ary, is easily obtained if one considers a family of $(m-1)$ increasing continuous functions $p_j:[0,y_{max}] \to [0,1]$, with $j = 0,1/(m-1),2/(m-1),...,1$ and set $p_0 = 1 - p_1 - ... - p_{m-1}$. The probabilities $p_j$ ($j = 0,...,m$) depend on the value of the variable $y_n$, i.e., they are *place dependent probabilities*. In a certain sense they represent the "randomly changing environment" where the synthesis take place due to the concentration of the transcription factor. Indeed, if the protein production $y$ is zero then the probability $p_0$ that transcription is switched OFF is equal to 1. Place dependent probabilities have been introduced in the work of Karlin [86,87] in the context of learning models and have played an important role in the theory of *iterated function systems* (IFS) [71,72,88,89] and fractals.

**Non-constant reaction processes.** By assuming $\alpha, \beta, \gamma, \delta$ to be constant in time our model focuses on the effects caused by the intermittency of gene activation / deactivation and neglects the noise coming form protein synthesis and mRNA and protein degradation. These effects can be taken into account by allowing $\alpha_n, \beta_n, \gamma_n, \delta_n$ to be time-dependent stationary stochastic processes with bounded values. However, when the reaction processes are independent from the state processes $\xi_n, x_n$ and $y_n$ one may consider the constant rate equations with the corresponding mean values $\langle\alpha\rangle, \langle\beta\rangle, \langle\gamma\rangle, \langle\delta\rangle$ which are time-independent due to stationarity, replacing the original reaction rate processes. This *constant rate approximation* of the original system by constant rate equations may be employed when the noise levels of the reaction processes are much smaller then the noise levels of the state processes.

**Continuous time version.** Instead of considering the discrete-time equation (1), we could have proceeded to develop our theory directly for the continuous-time case, along the same lines of [40,41]. However, this would require some mathematical background from theory of random dynamical systems. In fact, using the theory expounded in [38] it is possible to extend our results to the continuous-time case. Using the continuous-time version makes it easier to compute the first two moments of $x$ and $y$. In fact the mean values at equilibrium are given by exactly the same expressions (9)–(10) as in the discrete-time version, whereas the variances are given in terms of the covariances by $\sigma^2_x = \delta/\gamma \, \sigma_{\xi,x}$ and $\sigma^2_y = \beta/\alpha \, \sigma_{x,y}$. However, it is much harder to compute the covariances $\sigma_{\xi,x}$ and $\sigma_{x,y}$ in the continuous-time case since they are given by integral equations.

**Non-linear equations.** Another possible extension is the inclusion of *non-linear* terms in the equations (2)–(5). If the nonlinearity is small in comparison with the mean values of the rates appearing in the linear part then the qualitative features of the system is preserved. In other words, the model is *robust* under small perturbations, due the *hyperbolicity* of the linear part. The presence of nonlinearities would result in the loss of affine-preservation property of our model and one would have to work in the much more general class of smooth random dynamical systems. Nevertheless, it would open the door for the appearance of *stochastic bifurcations* [38,90]

**State-dependent reaction processes.** One interesting possibility is to allow the reaction rate parameters to depend on the state variables through input functions as well. In particular, linear input functions would reproduce a similar effect as the typical linear dependence on the reaction rates very commonly present in master equation approaches to gene expression [7,8].

**Differential and difference equations with random delay.** It has been suggested [12,91] that the delay between transcription and translation due to transport form the nucleus is an important factor for gene expression in eukaryotic cells. Indeed, it is possible to introduce *random delays* in our model in order to account for these factors by employing the theory of random dynamical systems with random delays [92,93]. Consider a sequence of random variables $\tau(n)$ assuming non-negative integer values representing the random delay times and replace the protein difference equation (3) by the delayed difference equation $y_{n+1} = \alpha y_n + \beta x_{n-\tau(n)}$. In the continuous time case, one may replace the corresponding protein difference equation by a delayed differential equation with time delay in the $x$ variable.

**Relation with the master equations approach.** It is possible to derive master equations from the continuous time version of the model discussed here. As a matter of fact, such master equations have been introduced in physics by Kubo [94,95], under the whimsical name of "stochastic Liouville equation" and have been "re-discovered" several times [96,97]. In biology, the same type of system has been "re-discovered" several times and has been called the "hybrid model for gene expression" going back to [61]. In [40–42] it is shown how to derive a master equation for the occupancy probabilities of a continuous time one-gene system described by an affine RDS and how to solve this equation in the case of a single gene. When the continuous-time state variables $x$ and $y$ are time-discretized, the master equation obtained is very similar to a birth and death process in a random environment [13], which is the same type of stochastic process used in the master equation approach to gene expression. A more thorough study of the hybrid model would be an important step in relating our formalism with the master equations approach to gene expression.

**Finite population effects and the Langevin approach.** It is possible to relate the model discussed here with the Langevin approach when accounting for fluctuations due to finite population effects. Consider a set of $N$ independent copies of the gene being considered such that the mean value $\langle \xi_n \rangle_N$ of the process $\xi_n$ over the ensemble evolves as $\langle \xi_n \rangle_N = \langle \xi \rangle + \sigma_\xi w_{N,n}$, where $\langle \xi \rangle$ is time average of the process $\xi_n$. Here, $w_{N,n}$ is a *discrete-time white noise process* with mean $\langle w_{N,n} \rangle = 0$, variance $\langle w_{N,n} w_{N,n} \rangle = 1$, covariance $\langle w_{N,n} w_{N,n-k} \rangle = 0$ for all $n$ and all $k \neq 0$ and $\lim_{N \to \infty} w_{N,n} = 0$. Thus, according to equation (2), the mean values of $x$ with respect to the $N$-ensemble evolve as $\langle x_{n+1} \rangle_N = (1 - \gamma)\langle x_n \rangle_N + \delta \langle \xi \rangle + \sigma_\xi w_{N,n}$, which is simply a discrete-time Langevin-like equation, commonly used in several studies of gene expression fluctuations with respect to cell populations. Therefore, the model presented here can be viewed as prototype gene whose mean behavior taken over an ensemble of $N$ independent copies may be described by a Langevin equation.

**Acknowledgments.** Some of the ideas presented in this paper were obtained in collaboration with F.R.A. Bosco who suddenly passed away in December of 2012, during this period he was supported by FAPESP (Brazil) through the grant 08/04531-2 (2009-2011). F. Antoneli was supported by CNPq (Brazil) through the grant PQ-306362/2012-0 and thanks Martin Golubitsky for the support while visiting the MBI – OSU (Mathematical Biosciences Institute – Ohio State University). R.C. Ferreira was supported by CNPq (Brazil). M.R.S. Briones is supported by FAPESP, CNPq (Brazil) and by an International Research Scholar grant from the Howard Hughes Medical Institute (USA).

# References

[1] Munsky B, Neuert G, van Oudenaarden A. Using gene expression noise to understand gene regulation. Science 2012;336:183–7. doi:10.1126/science.1216379.

[2] Maamar H, Raj A, Dubnau D. Noise in gene expression determines cell fate in Bacillus subtilis. Science 2007;317:526–9.

[3] Raser JM, O'Shea EK. Control of stochasticity in eukaryotic gene expression. Science 2004;304:1811–4. doi:10.1126/science.1098641.

[4] Raser JM, O'Shea EK. Noise in gene expression: origins, consequences, and control. Science 2005;309:2010–3. doi:10.1126/science.1105891.

[5] Elowitz MB, Levine AJ, Siggia ED, Swain PS. Stochastic gene expression in a single cell. Science 2002;297:1183–6. doi:10.1126/science.1070919.

[6] Orphanides G, Reinberg D. A unified theory of gene expression. Cell 2002;108:439–51.

[7]  Paulsson J. Summing up the noise in gene networks. Nature 2004;427:415–8. doi:10.1038/nature02257.

[8]  Paulsson J. Models of stochastic gene expression. Phys Life Rev 2005;2:157–75.

[9]  Ozbudak EM, Thattai M, Kurtser I, Grossman AD, van Oudenaarden A. Regulation of noise in the expression of a single gene. Nat Genet 2002;31:69–73.

[10] Berg J. Out-of-Equilibrium Dynamics of Gene Expression and the Jarzynski Equality. Phys Rev Lett 2008;100:188101–1(4).

[11] Lei J. Stochasticity in single gene expression with both intrinsic noise and fluctuation in kinetic parameters. J Theor Biol 2009;256:485–92. doi:10.1016/j.jtbi.2008.10.028.

[12] Kaern M, Elston TC, Blake WJ, Collins JJ. Stochasticity in Gene Expression: from Theories to Phenotypes. Nat Genet Rev 2005;6:451–62. doi:10.1038/nrg1615.

[13] Peccoud J, Ycart B. Markovian modeling of gene product synthesis. Theor Popul Biol 1995;48:222–34.

[14] Hornos JEM, Schultz D, Innocentini GCP, Wang J, Walczak AM, Onuchic JN, et al. Self-regulating gene: An exact solution. Phys Rev E 2005;72:e051907. doi:10.1103/PhysRevE.72.051907.

[15] Ramos AF, Hornos JEM. Symmetry and Stochastic Gene Regulation. Phys Rev Lett 2007;99:e108103. doi:10.1103/PhysRevLett.99.108103.

[16] Innocentini GCP, Hornos JEM. Modeling stochastic gene expression under repression. J Math Biol 2007;55:413–31. doi:10.1007/s00285-007-0090-x.

[17] Ramos AF, Innocentini GCP, Forger M, Hornos JEM. Symmetry in biology: from genetic code to stochastic gene regulation. IET Syst Biol 2010;4:311–29. doi:10.1049/iet-syb.2010.0058.

[18] Ramos AF, Innocentini GCP, Hornos JEM. Exact time-dependent solutions for a self-regulating gene. Phys Rev E 2011;83:e062902. doi:10.1103/PhysRevE.83.062902.

[19] Radulescu O, Innocentini GCP, Hornos JEM. Relating network rigidity, time scale hierarchies, and expression noise in gene networks. Phys Rev E 2012;85:e041919. doi:10.1103/PhysRevE.85.041919.

[20] Innocentini GCP, Forger M, Ramos AF, Radulescu O, Hornos JEM. Multimodality and Flexibility of Stochastic Gene Expression. Bull Math Biol 2013;75:2600–30. doi:10.1007/s11538-013-9909-3.

[21] Sasai M, Wolynes PG. Stochastic gene expression as a many body problem. PNAS 2003;100:2374–9.

[22] Gillespie DT. Exact stochastic simulation of coupled chemical reactions. J Phys Chem 1977;81:2340–61. doi:10.1021/j100540a008.

[23] Gillespie DT. Stochastic simulation of chemical kinetics. Annu Rev Phys Chem 2007;58:35–55. doi:10.1146/annurev.physchem.58.032806.104637.

[24] Kuramoto Y. Chemical Oscillations, Waves, and Turbulence. Dover edition 2003. New York: Dover Publications; 1984.

[25] Watts D, Strogatz S. Collective dynamics of small-world networks. Nature 1998;393:440–2.

[26] Albert R, Barabási L. Statistical mechanics of complex networks. Rev Mod Phys 2002;74:47–97.

[27] Newman MEJ. The structure and function of complex networks. SIAM Rev 2003;45:167–256.

[28] Stewart I, Golubitsky M, Pivato M. Symmetry groupoids and patterns of synchrony in coupled cell networks. SIAM J Appl Dyn Syst 2003;2:609–46.

[29] Golubitsky M, Stewart I, Török A. Patterns of synchrony in coupled cell networks with multiple arrows. SIAM J Appl Dyn Syst 2005;4:78–100.

[30] Field M. Combinatorial dynamics. Dyn Syst 2004;19:217–43.

[31] Golubitsky M, Stewart I. Nonlinear Dynamics of Networks: The groupoid formalism. Bull AMS 2006;43:305–64.

[32] Aguiar M, Ashwin P, Dias A, Field M. Dynamics of Coupled Cell Networks: Synchrony, Heteroclinic Cycles and Inflation. J Nonlinear Sci 2011;21:271–323. doi:10.1007/s00332-010-9083-9.

[33] Antoneli F, Dias APS, Golubitsky M, Wang Y. Patterns of synchrony in lattice dynamical systems. Nonlinearity 2005;18:2193. doi:10.1088/0951-7715/18/5/016.

[34] Antoneli F, Dias APS, Pinto CMA. Quasi-periodic states in coupled rings of cells. Commun Nonlinear Sci Numer Simul 2010;15:1048–62. doi:10.1016/j.cnsns.2009.05.035.

[35] Lee TI, Rinaldi NJ, Robert F, Odom DT, Bar-Joseph Z, Gerber GK, et al. Transcriptional Regulatory Networks in Saccharomyces cerevisiae. Science 2002;298:799–804. doi:10.1126/science.1075090.

[36] Ferreira RC, Bosco FAR, Paiva PB, Briones MRS. Minimization of transcriptional temporal noise and scale invariance in the yeast genome. Genet Mol Res 2007;6:397–411.

[37] Kifer Y. Ergodic Theory of Random Transforms. vol. 10. Boston: Birkhäuser; 1986.

[38] Arnold L. Random dynamical systems. Berlin: Springer-Verlag; 1998.

[39] Karlin S, Taylor HM. A First Course in Stochastic Processes. Academic Press; 1975.

[40] Lipniacki T, Paszek P, Marciniak-Czochra A, Brasier AR, Kimmel M. Transcriptional stochasticity in gene expression. J Theor Biol 2006;238:348–67. doi:10.1016/j.jtbi.2005.05.032.

[41] Bobrowski A, Lipniacki T, Pichór K, Rudnicki R. Asymptotic behavior of distributions of mRNA and protein levels in a model of stochastic gene expression. J Math Anal Appl 2007;333:753–69. doi:10.1016/j.jmaa.2006.11.043.

[42] Smiley MW, Proulx SR. Gene expression dynamics in randomly varying environments. J Math Biol 2010;61:231–51. doi:10.1007/s00285-009-0298-z.

[43] Mackey MC, Tyran-Kamińska M, Yvinec R. Molecular distributions in gene regulatory dynamics. J Theor Biol 2011;274:84–96. doi:10.1016/j.jtbi.2011.01.020.

[44] Santillán M. On the Use of the Hill Functions in Mathematical Models of Gene Regulatory Networks. Math Model Nat Phenom 2008;3:85–97. doi:10.1051/mmnp:2008056.

[45] Arnold L, Crauel H. Iterated function systems and multiplicative ergodic theory. Diffus. Process. Relat. Probl. Anal. Vol II Charlotte NC 1990, vol. 27, Boston, MA: Birkhäuser; 1992, p. 283–305.

[46] Alon U. An Introduction to Systems Biology: Design Principles of Biological Circuits. Chapman & Hall/CRC; 2007.

[47] Crauel H, Flandoli F. Additive Noise Destroys a Pitchfork Bifurcation. J Dyn Differ Equ 1998;10:259–74. doi:10.1023/A:1022665916629.

[48] Callaway M, Doan TS, Lamb JSW, Rasmussen M. The dichotomy spectrum for random dynamical systems and pitchfork bifurcations with additive noise. ArXiv13106166 Math 2013.

[49] Thattai M, van Oudenaarden A. Intrinsic noise in gene regulatory networks. PNAS 2001;98:8614–9.

[50] Tyson JJ, Novák B. Functional Motifs in Biochemical Reaction Networks. Annu Rev Phys Chem 2010;61:219–40. doi:10.1146/annurev.physchem.012809.103457.

[51] Barnsley MF. SuperFractals. 1 edition. Cambridge ; New York: Cambridge University Press; 2006.

[52] Ermentrout B. XPPAUT version 6.1. 2012;http://www.math.pitt.edu/~bard/xpp/xpp.html.

[53] Ermentrout B. Simulating, analyzing, and animating dynamical systems a guide to XPPAUT for researchers and students. Philadelphia, Pa.: Society for Industrial and Applied Mathematics (SIAM, 3600 Market Street, Floor 6, Philadelphia, PA 19104); 2002.

[54] R Core Team. R: A language and environment for statistical computing. 2015;http://www.R-project.org/.

[55] McAdams HH, Arkin A. Stochastic mechanisms in gene expression. PNAS 1997;94:814–9.

[56] Arkin A, Ross J, McAdams HH. Stochastic Kinetic Analysis of Developmental Pathway Bifurcation in Phage λ-Infected Escherichia coli Cells. Genetics 1998;149:1633–48.

[57] Gilman A, Arkin AP. GENETIC "CODE": Representations and Dynamical Models of Genetic Components and Networks. Annu Rev Genomics Hum Genet 2002;3:341–69. doi:10.1146/annurev.genom.3.030502.111004.

[58] Ackers GK, Johnson AD, Shea MA. Quantitative model for gene regulation by lambda phage repressor. Proc Natl Acad Sci U S A 1982;79:1129–33.

[59] Shea MA, Ackers GK. The OR control system of bacteriophage lambda. A physical-chemical model for gene regulation. J Mol Biol 1985;181:211–30.

[60] Ko MSH. A stochastic model for gene induction. J Theor Biol 1991;153:181–94. doi:10.1016/S0022-5193(05)80421-7.

[61] Kepler TB, Elston TC. Stochasticity in transcriptional regulation: origins, consequences, and mathematical representations. Biophys J 2001;81:3116–36. doi:10.1016/S0006-3495(01)75949-8.

[62] Pirone JR, Elston TC. Fluctuations in transcription factor binding can explain the graded and binary responses observed in inducible gene expression. J Theor Biol 2004;226:111–21.

[63] Simpson ML, Cox CD, Sayler GS. Frequency domain chemical Langevin analysis of stochasticity in gene transcriptional regulation. J Theor Biol 2004;229:383–94. doi:10.1016/j.jtbi.2004.04.017.

[64] Takasuka N, White MR, Wood CD, Robertson WR, Davis JR. Dynamic changes in prolactin promoter activation in individual living lactotrophic cells. Endocrinology 1998;139:1361–8. doi:10.1210/endo.139.3.5826.

[65] Stirland JA, Seymour ZC, Windeatt S, Norris AJ, Stanley P, Castro MG, et al. Real-time imaging of gene promoter activity using an adenoviral reporter construct demonstrates transcriptional dynamics in normal anterior pituitary cells. J Endocrinol 2003;178:61–9.

[66] Quas A, Bose C, Bahsoun W. Deterministic representation for position dependent random maps. Discrete Contin Dyn Syst 2008;22:529–40. doi:10.3934/dcds.2008.22.529.

[67] Kwiecińska AA, Słomczyński W. Random dynamical systems arising from iterated function systems with place-dependent probabilities. Stat Probab Lett 2000;50:401–7.

[68] Lasota A, Mackey MC. Chaos, fractals, and noise: Stochastic aspects of dynamics. Berlin: Springer-Verlag; 1993.

[69] Jaroszewska J. Iterated function systems with continuous place dependent probabilities. Univ Iagell Acta Math 2002;40:137–46.

[70] Stenflo Ö. Ergodic theorems for time-dependent random iteration of functions. Fractals Valletta 1998, River Edge, NJ: World Scientific; 1998, p. 129–36.

[71] Barnsley MF, Demko SG, Elton JH, Geronimo JS. Invariant measures for Markov processes arising from iterated function systems with place-dependent probabilities. Ann Inst Henri Poincaré Probab Stat 1988;24:367–94.

[72] Barnsley MF, Demko SG, Elton JH, Geronimo JS. Erratum: "Invariant measures for Markov processes arising from iterated function systems with place-dependent probabilities. Ann. Inst. H. Poincaré Probab. Statist. 24(3):367–394." Ann Inst H Poincaré Probab Stat 1988;25:589–90.

[73] Buse O, Pérez R, Kuznetsov A. Dynamical properties of the repressilator model. Phys Rev E 2010;81:066206. doi:10.1103/PhysRevE.81.066206.

[74] Müller S, Hofbauer J, Endler L, Flamm C, Widder S, Schuster P. A generalized model of the repressilator. J Math Biol 2006;53:905–37. doi:10.1007/s00285-006-0035-9.

[75] Ferreira RC, Bosco FAR, Briones MRS. Scaling properties of transcription profiles in gene networks. Int J Bioinform Res Appl 2009;5:178–86.

[76] Innocentini GCP, Forger M, Radulescu O, Antoneli F. Protein Synthesis Driven by Dynamical Stochastic Transcription. Bull Math Biol 2015:1–22. doi:10.1007/s11538-015-0131-3.

[77] Arnold L, Crauel H. Random dynamical systems. In: Arnold L, Crauel H, Eckmann J-P, editors. vol. 1486, Berlin: Springer-Verlag; 1991, p. 1–22.

[78] Arnold L. Random dynamical systems. In: Arnold L, Crauel H, Eckmann J-P, editors. vol. 1609, Berlin: Springer-Verlag; 1995, p. 1–43.

[79] Kifer Y, Liu P-D. Random Dynamics. Handb. Dyn. Syst. Vol 1B, Amsterdam: Elsevier; 2006, p. 379–499.

[80] Hutchinson JE. Fractals and self-similarity. Indiana Univ Math J 1981;30:713–47.

[81] Solomyak B. Notes on Bernoulli convolutions. Fractal Geom. Appl. Jubil. Benoı̃ T Mand. Part 1, vol. 72, Providence, RI: Amer. Math. Soc.; 2004, p. 207–30.

[82] Erdös P. On a family of symmetric Bernoulli convolutions. Amer J Math 1939;61:974–6.

[83] Solomyak B. On the Random Series $\sum\pm\lambda^n$ (an Erdös Problem). Ann Math 1995;142:611–25.

[84] Shmerkin P. Overlapping self-affine sets. Indiana Univ Math J 2006;55:1291–331. doi:10.1512/iumj.2006.55.2718.

[85] Peres Y, Schlag W, Solomyak B. Sixty years of Bernoulli convolutions. In: Bandt C, Graf S, Zähle M, editors. Fractal Geom. Stoch. II, vol. 46, Berlin: Birkhäuser; 2000, p. 36–65.

[86] Karlin S. Some random walks arising in learning models I. Pac J Math 1953;3:725–56.

[87] Stenflo Ö. A note on a theorem of Karlin. Stat Probab Lett 2001;54:183–7.

[88] Barnsley MF, Elton JH, Hardin DP. Recurrent iterated function systems. Constr Approx Int J Approx Expans 1989;5:3–31.

[89] Elton JH. An ergodic theorem for iterated maps. Ergod Theory Dyn Syst 1987;7:481–8.

[90] Arnold L, Boxler P. Additive noise turns a hyperbolic fixed point into a stationary solution. In: Arnold L, Crauel H, Eckmann J-P, editors. vol. 1486, Berlin: Springer-Verlag; 1991, p. 159–64. doi:10.1007/BFb0086665.

[91] Kerszberg M. Noise, delays, robustness, canalization and all that. Curr Opin Genet Dev 2004;14:440–5.

[92] Crauel H, Doan TS, Siegmund S. Difference equations with random delay. J Differ Equ Appl 2009;15:627–47. doi:10.1080/10236190802612865.

[93] Siegmund S, Doan TS. Differential Equations with Random Delay. In: Mallet-Paret J, Wu J, Yi Y, Zhu H, editors. Infin. Dimens. Dyn. Syst., vol. 64, New York, NY: Springer New York; 2012, p. 279–303.

[94] Kubo R. Stochastic Liouville Equations. J Math Phys 1963;4:174–83. doi:10.1063/1.1703941.

[95] Kubo R. A Stochastic Theory of Line Shape. In: Shuler KE, editor. Adv. Chem. Phys., John Wiley & Sons, Inc.; 1969, p. 101–27.

[96] van Kampen NG. Stochastic processes in physics and chemistry. Amsterdam; Boston; London: Elsevier; 2007.

[97] van Kampen NG. Remarks on Non-Markov Processes. Braz J Phys 1998;28:90–6. doi:10.1590/S0103-97331998000200003.